\documentclass[12pt]{article}
\pdfoutput=1
\usepackage{titling}
\usepackage{amsmath}
\usepackage{slashed}
\usepackage{amssymb}
\usepackage{epsfig}
\usepackage{graphicx}
\usepackage{multirow}
\usepackage{color}
\usepackage[normal]{subfigure}
\usepackage{rotating}
\usepackage{hyperref}
\usepackage[margin=0.9in]{geometry}
\usepackage[table]{xcolor}
\usepackage{enumitem}
\usepackage[utf8x]{inputenc}
\usepackage[compress,numbers,sort]{natbib}
\usepackage{authblk}
\usepackage{colortbl}
\usepackage{pdflscape}
\usepackage{color}

\definecolor{nicered}{rgb}{0.6,0.1,0.1}
\definecolor{nicegreen}{rgb}{0.1,0.5,0.1}
\definecolor{niceblue}{rgb}{0,0,0.8}
\definecolor{mediumcandyapplered}{rgb}{0.99, 0.12, 0.07}
\definecolor{red}{rgb}{1.0, 0, 0}
\hypersetup{colorlinks,citecolor= nicegreen,linkcolor= nicered,urlcolor=niceblue}

\def\eq#1{{Eq.~(\ref{#1})}}
\def\eqs#1#2{{Eqs.~(\ref{#1})--(\ref{#2})}}
\def\fig#1{{Fig.~\ref{#1}}}
\def\figs#1#2{{Figs.~\ref{#1}--\ref{#2}}}
\def\Table#1{{Table~\ref{#1}}}

\def\sect#1{{Sect.~\ref{#1}}}
\def\sects#1#2{{Sects.~\ref{#1}--\ref{#2}}}
\def\app#1{{Appendix~\ref{#1}}}

\renewcommand{\bar}{\overline}

\definecolor{LightCyan}{rgb}{0.88,1,1}
\definecolor{piggypink}{rgb}{0.99, 0.87, 0.9}
\definecolor{applegreen}{rgb}{0.55, 0.71, 0.0}
\definecolor{darkpastelgreen}{rgb}{0.01, 0.75, 0.24}
\definecolor{green-yellow}{rgb}{0.68, 1.0, 0.18}
\definecolor{green-yellow2}{rgb}{0.78, 1.0, 0.68}

\newcommand{\beq}{\begin{equation}}
\newcommand{\eeq}{\end{equation}}
\newcommand{\bea}{\begin{eqnarray}}
\newcommand{\eea}{\end{eqnarray}}

\newcommand{\published}[1]{%
\gdef\puB{#1}}
\newcommand{\puB}{}

\title{\bf{Probing new electroweak states via precision measurements at the LHC
and future colliders
}}

\author[1,2]
{Luca Di Luzio\thanks{lukaluz@gmail.com}}
\author[3,2]
{Ramona Gr\"{o}ber\thanks{ramona.groeber@physik.hu-berlin.de}}
\author[4,5]
{Giuliano Panico\thanks{gpanico@ifae.es}}

\affil[1]{\emph{\normalsize Dipartimento di Fisica dell'Universit\`{a} di Pisa and INFN, Italy}}

\affil[2]{\emph{\normalsize Institute for Particle Physics Phenomenology, Department of Physics, 
Durham University, DH1 3LE, Durham, United Kingdom}}

\affil[3]{\emph{\normalsize Humboldt-Universit\"at zu Berlin, Institut f\"ur Physik, Newtonstr.~15, 
12489 Berlin, Germany}}

\affil[4]{\emph{\normalsize Deutsches Elektronen-Synchrotron (DESY), Notkestr.~85, 22607 Hamburg, Germany}}

\affil[5]{\emph{\normalsize IFAE and BIST, Universitat Aut\'onoma de Barcelona,
E-08193 Bellaterra, Barcelona, Spain}}

\date{}

\published{\flushright \vskip-0.5cm HU-EP-18/30, DESY 18-184}

\begin{document}

\maketitle
\begin{abstract}
\normalsize
Several new physics scenarios, motivated e.g.~by dark matter, feature new electroweakly charged states 
where the lightest particle in the multiplet is stable and neutral. In such cases direct searches at LHC are notoriously 
difficult, while electroweak precision tests both at hadron and lepton colliders offer the possibility to indirectly probe 
those states. In this work, we assess the sensitivity of the high-luminosity phase of the LHC on new electroweak multiplets 
via the modification of neutral and charged Drell-Yan processes, 
and compare the reach of future hadron and lepton colliders presently under consideration. 

\end{abstract}

\clearpage

\tableofcontents

\clearpage

\section{Introduction}

Massive particles carrying electroweak (EW) quantum numbers are predicted in many motivated extensions 
of the standard model (SM). For instance, neutralinos and charginos in supersymmetry (SUSY) 
or vector-like leptons in composite Higgs models.
Being only charged under ${\rm SU}(2)_L \times {\rm U}(1)_y$, 
they remain more elusive at the LHC with respect to coloured states, with current bounds well below the TeV scale. 
EW multiplets are particularly difficult to be probed in collider experiments 
whenever the lightest particle of the ${\rm SU}(2)_L$ $n$-plet is electrically neutral and stable. 
The current bounds, stemming  from disappearing track and mono-X searches are 
of the order of 100 -- 500 GeV (depending on the dimensionality $n$). 
On the other hand, this latter case is also of particular interest since it renders the lightest particle in the  $n$-plet
an ideal candidate for dark matter (DM).  To this end,  alternative ways of probing EW multiplets
are particularly welcome. 

In this paper we explore the possibility of indirectly probing new EW states 
through the precise measurements of neutral and charged Drell-Yan processes: 
$pp \to \ell^+ \ell^-$ and $pp\to \ell \nu$. In Ref.~\cite{Farina:2016rws} 
it was shown that the latter processes can be exploited at the LHC in order to  
perform accurate tests of the EW sector,  
by benefitting from the growth in energy of the corrections due to new physics (see also \cite{Cirigliano:2012ab,deBlas:2013qqa}). 
In Ref.~\cite{Farina:2016rws} the limit in which new physics is significantly heavier than the available collider energy was
considered.
In this case, if new physics affects mainly the gauge boson self energies, the corrections can be encoded
in the ``oblique'' parameters, namely the original Peskin-Takeuchi $S$, $T$ and $U$~\cite{Peskin:1991sw}, 
extended by the additional parameters $W$ and $Y$~\cite{Barbieri:2004qk}
which include higher-order terms in the momentum expansion 
(for an updated EW fit employing low-energy observables 
see e.g.~Ref.~\cite{Falkowski:2017pss}). 
On the other hand, the $W$ and $Y$ parameters induce corrections that grow in energy, therefore they can be easily accessed at the LHC
by exploiting the high reach in invariant mass $m_{\ell\ell}$ or transverse mass $m_T$. This allows to significantly
improve the sensitivity with respect to LEP.
On the contrary $S$, $T$ and $U$ give an overall rescaling of the cross section, thus they are dominantly tested at
the $Z$ pole. In this case LHC can hardly compete with LEP.\footnote{In Refs.~\cite{Panico:2017frx,Franceschini:2017xkh, Banerjee:2018bio} it was shown that also precision measurements of di-boson production can provide further tests of the SM effective theory. Exploitation of these channels allows LHC to reach a sensitivity comparable to the LEP one also for
observables analogous to the $S$ parameter. Further indirect tests of dark fermions or scalars can come from Higgs coupling measurements \cite{Voigt:2017vfz}.}

When the new physics is very light, much below the effective collider energy, the simplified
description in terms of oblique parameters is not appropriate anymore.
In this case the effects of light new physics can be seen as a modification of the running of the EW gauge couplings due to the
change in the SM beta functions for scales $\mu > m$, where $m$ generically denotes the mass scale of the new state. 
The regime of very light new physics has been analysed e.g.~in Refs.~\cite{Becciolini:2014lya,Alves:2014cda,Gross:2016ioi}. 

Since the current bounds on EW multiplets are of the order of few $100$ GeV we are mostly interested in an 
intermediate regime where none of the two limits above can be applied. 
We will instead  keep the full corrections due to the new state in the gauge boson propagators. 
Such an analysis has been done  previously in Ref.~\cite{Harigaya:2015yaa} in the context of future $e^+e^-$ lepton 
colliders 
and in Ref.~\cite{Matsumoto:2017vfu} for LHC and its high-luminosity phase (HL-LHC). 
While Ref.~\cite{Matsumoto:2017vfu} considers only the neutral Drell-Yan production, 
we include in our analysis also the charged Drell-Yan process and show that it actually leads 
to stronger bounds than the neutral channel. 
We pay special attention to the treatment of uncertainties and show where an improvement 
of systematic errors and/or parton distribution functions (PDF) can lead to a substantial improvement of the 
bound. Furthermore, we analyse the sensitivity of future facilities presently under discussion, 
such as a 28 TeV high-energy LHC (HE-LHC) \cite{Benedikt:2018ofy} and a 100 TeV future circular collider (FCC-100) \cite{Hinchliffe:2015qma, Golling:2016gvc}, as well as 
high-energy lepton colliders including $e^+e^-$ machines 
like the Compact Linear Collider (CLIC) \cite{Linssen:2012hp} and muon colliders, with the multi-TeV options MAP (muon accelerator program) \cite{Delahaye:2013jla} based on proton scattering on a target and LEMMA (low emittance muon accelerator) \cite{Antonelli:2015nla, Collamati:2017jww, Boscolo:2018tlu} based on positron scattering on a target.

The paper is structured as follows. In \sect{physcaseew} we introduce the physics case for EW multiplets. 
In \sect{sec:propagators} we describe the parametrization of the new physics corrections to the gauge boson propagators. 
In \sect{sec:HL-LHC} we present the analysis for the HL-LHC and future hadron colliders 
which represents the central part of our work, while 
in \sect{ewptCLIC} we study the sensitivity of future lepton collider options.  
Finally in \sect{compDS} we briefly compare our bounds on EW multiplets 
with direct searches and conclude in \sect{sec:conclusion}. Appendices~\ref{App:pert} and \ref{App:additional}
are devoted to  the collection of some additional results and technical details.

\section{Physics case for new EW multiplets}
\label{physcaseew}

New EW states charged under ${\rm SU}(2)_L \times {\rm U}(1)_y$, which are generically denoted by their
quantum numbers $\chi \sim (1,n,y)$, with the three entries denoting the 
${\rm SU}(3)_c \times {\rm SU}(2)_L \times {\rm U}(1)_y$ representation, 
appear in many motivated beyond-the-SM scenarios. 
The EW sector of SUSY comprising the wino/higgsino system is certainly one of the most compelling cases. 
Larger multiplets with $n > 3$ can also be motivated by DM, 
if the lightest particle in the $n$-dimensional multiplet is stable and neutral. 
In the following, we briefly review a few frameworks which 
motivate the existence of large EW multiplets from the standpoint of 
accidental global symmetries.  

\subsection{Minimal (milli-charged) Dark Matter} 

The idea behind Minimal Dark Matter (MDM) \cite{Cirelli:2005uq,Cirelli:2007xd,Cirelli:2009uv} 
is to introduce a single EW multiplet $\chi$ 
which is accidentally stable at the renormalizable level due to the SM gauge symmetry. 
One further assumes $y=0$ (to avoid direct detection bounds from $Z$ exchange) 
and that the lightest particle in the multiplet is neutral. The latter is actually an automatic feature if the mass splitting 
within the $n$-plet
is purely radiative as in the case of fermions with $n>3$. On the contrary, scalars can receive a 
model-dependent tree-level splitting from the scalar potential, 
which we assume to be subleading.
The contribution to the relic density is completely fixed 
by the EW gauge interactions and the mass of the new state $m_\chi$, 
thus making the framework extremely predictive.  

If one further requires that the 
theory remains weakly coupled up to the Planck scale and that the gauge quantum numbers of 
$\chi$ are such that no operators with dimension smaller than $6$
can mediate the decay of $\chi$,\footnote{Operators with dimension $\leq 5$ would lead to 
a too fast $\chi$ decay, even with a Planck scale cutoff.}
only one multiplet is allowed, namely the Majorana fermion representation $(1,5,0)_{\rm MF}$.\footnote{Originally also the 
real scalar representation $(1,7,0)_{\rm RS}$ was included in the list, 
but it was shown later in Ref.~\cite{DiLuzio:2015oha} that a previously overlooked $d=5$ operator leads to 
a loop-induced decay of $\chi$, 
with a lifetime shorter than the age of the Universe.} 
To be completely general, in the following, we will however consider multiplets of different kind,
namely in the real scalar, complex scalar, Majorana fermion, Dirac fermion representations, which
we denote by the labels RS, CS, MF, DF respectively.

The MDM framework was extended in Ref.~\cite{DelNobile:2015bqo} to contemplate the possibility of a milli-charge 
$\epsilon \ll 1$. Bounds from DM direct detection imply $\epsilon \lesssim 10^{-9}$. 
The milli-charge has hence no bearings for collider phenomenology, but it ensures the (exact) stability of the 
lightest particle in the EW multiplet 
due to the SM gauge symmetry, in the same spirit of the original MDM formulation.  
A notable feature of the milli-charged scenario is that the contribution of the complex 
multiplet to the relic density gets doubled compared to 
the case of a single real component 
(thus making the thermal mass roughly a factor $\sqrt{2}$ smaller). On the other hand, the 
number of degrees of freedom are also doubled, 
thus improving the indirect testability of those scenarios via EW precision tests at colliders. 

The MDM candidates (including for completeness also the higgsino-like $(1,2,1/2)_{\rm DF}$ and 
wino-like $(1,3,0)_{\rm MF}$ DM, which require a stabilization mechanism beyond the SM gauge symmetry) 
are summarized in \Table{tableMDM}, together with their thermal mass saturating the 
DM relic density\footnote{The thermal masses in the $\epsilon = 0$ cases are extracted 
from Ref.~\cite{Mitridate:2017izz} which takes into account both Sommerfeld enhancement 
and bound state formation effects. 
In the cases $\epsilon \neq 0$ 
we quote instead the results from Ref.~\cite{DelNobile:2015bqo}, which however 
do not include effects from bound state formation that are expected to sizeable  
for $n\gtrsim 5$ (e.g.~in the case of $(1,5,0)_{\rm MF}$ the 
inclusion of bound state effects leads to a $20\%$ increase of the thermal mass \cite{Mitridate:2017izz}).}
and the projected $95 \%$ confidence level (CL) 
exclusion limits of five representative future colliders: 
HL-LHC ($\sqrt{s} = 14$ TeV and $L=3$/ab), 
HE-LHC ($\sqrt{s} = 28$ TeV and $L=10$/ab), 
FCC-100 ($\sqrt{s} = 100$ TeV and $L=20$/ab), 
CLIC-3 ($\sqrt{s} = 3$ TeV and $L=4/\text{ab}$), 
Muon-14 ($\sqrt{s} = 14$ TeV and $L=20$/ab).
The details of the analysis will be presented in \sects{sec:HL-LHC}{ewptCLIC}.
\begin{table}[tp]
\centering
\begin{tabular}{@{} |l|c|c|c|c|c|c| @{}}
\hline
$\chi$ / $m_{\chi}$\,[TeV] & DM
& HL-LHC 
& HE-LHC
& FCC-100
& CLIC-3 
& Muon-14 \\
\hline
\hline
$(1,2,1/2)_{\rm DF}$ & 1.1 & -- & -- & -- & 0.4 & 0.6 \\ 
$(1,3,\epsilon)_{\rm CS}$ & 1.6 & -- & -- & -- & 0.2 & 0.2 \\ 
$(1,3,\epsilon)_{\rm DF}$ & 2.0 & -- & 0.6 & 1.5 & \cellcolor{piggypink}0.8 $\&$ [1.0, 2.0] 
& \cellcolor{piggypink}2.2 $\&$ [6.3, 7.1]  \\ 
$(1,3,0)_{\rm MF}$ & 2.8 & -- & -- & 0.4 & 0.6 $\&$ [1.2, 1.6] & 1.0 \\ 
$(1,5,\epsilon)_{\rm CS}$ & 6.6 & 0.2 & 0.4 & 1.0 & 0.5 $\&$ [0.7,1.6] & 1.6 \\ 
$(1,5,\epsilon)_{\rm DF}$ & 6.6 & 1.5 & 2.8 & \cellcolor{piggypink}7.1 & 3.9 & \cellcolor{piggypink}11 \\ 
$(1,5,0)_{\rm MF}$ & 14 & 0.9 & 1.8 & 4.4 & 2.9 & 3.5 $\&$ [5.1, 8.7] \\ 
$(1,7,\epsilon)_{\rm CS}$ & 16 & 0.6 & 1.3 & 3.2 & 2.4 & 2.5 $\&$ [3.5, 7.4] \\ 
$(1,7,\epsilon)_{\rm DF}$ & 16 & 2.1 & 4.0 & 11 & 6.4 & \cellcolor{piggypink}18 \\ 
  \hline
\end{tabular}
\caption{\it Pure higgsino/wino-like DM and MDM candidates, 
together with the corresponding masses saturating the DM relic density (second column) 
and the projected $95 \%$ CL exclusion limits from EW precision tests at 
HL-LHC, 
HE-LHC, 
FCC-100, 
CLIC-3 
and Muon-14 (see text for details about center-of-mass energies and luminosities).  
In the last two columns the numbers in square brackets stand for a mass interval exclusion. 
The cases where the DM hypothesis could be fully tested are emphasized in light red. 
}\label{tableMDM}
\end{table}

We can anticipate here some results of our analysis. The HL-LHC and the HE-LHC are not able to test
any of the DM candidates for masses which allow these multiplets to saturate the whole DM relic density.
The FCC-100, on the other hand, could fully test the $(1,5,\epsilon)_{\rm DF}$ 
candidate and would come close
to test the interesting mass range for the $(1,3,\epsilon)_{\rm DF}$ 
and $(1,7,\epsilon)_{\rm DF}$ multiplets.
Lepton colliders are usually better at testing small multiplets, which are difficult to probe at hadron colliders.
CLIC-3 and Muon-14 could fully test the $(1,3,\epsilon)_{\rm DF}$ multiplet. Muon-14 would also surpass
the FCC-100 sensitivity on both the $(1,5,\epsilon)_{\rm DF}$ and the $(1,7,\epsilon)_{\rm DF}$ multiplets,
reaching the masses that saturate the DM relic density. It could also test a significant fraction of the mass range for the
$(1,5,0)_{\rm MF}$ multiplet.

\subsection{Accidental Matter} 

From a more phenomenological point of view, larger multiplets with $n>3$ 
are also motivated by the fact that 
they automatically 
respect the accidental and approximate symmetry structure of the SM and are hence screened 
from low-energy probes such as baryon and lepton number and flavour/CP violating processes.
Particularly interesting multiplets are the ones that satisfy the following requirements~\cite{DiLuzio:2015oha}:
$i)$ automatically preserve the accidental and approximate symmetry structure of the SM; 
$ii)$ are cosmologically viable; 
$iii)$ form consistent effective field theories (EFTs)
with a cut-off scale as high as $10^{15}$ GeV (as suggested by neutrino masses).
These multiplets are easily compatible with the stringent flavor tests and typically present peculiar collider
signatures, for instance long-lived particles.
A finite list of multiplets satisfy the above criteria \cite{DiLuzio:2015oha}, 
and among those a subset feature a neutral lightest state:
$(1,5,0)_{\rm RS}$, $(1,5,1)_{\rm CS}$, $(1,5,2)_{\rm CS}$, $(1,7,0)_{\rm RS}$, 
$(1,4,3/2)_{\rm DF}$, $(1,5,0)_{\rm MF}$. 
These multiplets are hence a natural target for our study.
It turns out that the value of the hypercharge, unless exotically large, 
plays a subleading role for the extraction of the bound.\footnote{E.g.~in the case 
of $(1,5,2)_{\rm CS}$ the bound on the mass gets strengthen by $5\%$ 
compared to $(1,5,0)_{\rm CS}$.} 
Hence, instead of reporting explicitly 
the projected reach for all the accidental matter candidates, we refer directly to the results in 
\sects{sec:HL-LHC}{ewptCLIC}.

\section{Universal EW corrections to $2 \to 2$ fermion processes}\label{sec:propagators}

In this work we consider the universal EW corrections to $2 \to 2$ processes involving SM fermions in the final state,  
stemming from a new scalar/fermion multiplet $\chi \sim (1,n,y)$. 
We further assume that $i)$ $\chi$ does not interact at the renormalizable level with 
the SM matter fields and $ii)$ the mass splitting within the $n$-plet is negligible. 
While the former assumption ensures that the radiative corrections can be  
encoded into the universal modification of the EW gauge boson propagators, 
the latter one is a simplification which 
becomes asymptotically good 
in the regime $m_\chi \gg m_Z$, relevant for future colliders.     
It is important to stress, however, that the above assumptions are automatically satisfied for 
fermions with $n>3$, while in the case of scalars they 
further require that interaction terms with the SM Higgs are subleading. 

The most useful observables at hadron colliders turn out to be the neutral and charged DY processes 
with leptons $\ell = e,\mu$ in the final states: 
$pp \to \ell^+\ell^-$ and $pp \to \ell\nu$, while at lepton colliders one can consider the neutral current 
processes $\ell^+\ell^- \to f \bar{f}$ (with $f$ denoting a SM fermion). 
In the following, we describe the formalism for deriving 
the modified EW gauge boson propagators, which is common to all 
$2 \to 2$ fermion processes. For related analyses see also Refs.~\cite{Harigaya:2015yaa,Matsumoto:2017vfu}. 

\subsection{Form factors}\label{sec:form_factors}

The modifications of the EW gauge boson propagators due to the new state $\chi \sim (1,n,y)$ 
is parametrized via the inclusion of the following form factors in the effective Lagrangian 
\beq 
\label{LEWIMPs}
\mathcal{L}_{\rm eff} = \mathcal{L}_{\rm SM} 
+ \frac{g^2 C^{\rm eff}_{WW}}{8} W^a_{\mu\nu} \Pi(-D^2/m_{\chi}^2) W^{a\mu\nu} 
+ \frac{g'^2 C^{\rm eff}_{BB}}{8} B_{\mu\nu} \Pi(-\partial^2/m_{\chi}^2) B^{\mu\nu} \, , 
\eeq
where 
\begin{equation}\label{eq:def_C}
C^{\rm eff}_{WW} = \kappa (n^3-n)/6\,, 
\qquad \quad
C^{\rm eff}_{BB} = 2 \kappa n y^2\,,
\end{equation} 
and $\kappa = 1/2,1,4,8$, respectively for $\chi$ being 
a real scalar (RS), complex scalar (CS), Majorana fermion (MF), Dirac fermion (DF). 

\begin{figure}[t]
\begin{center}
\includegraphics[width=0.5 \textwidth]{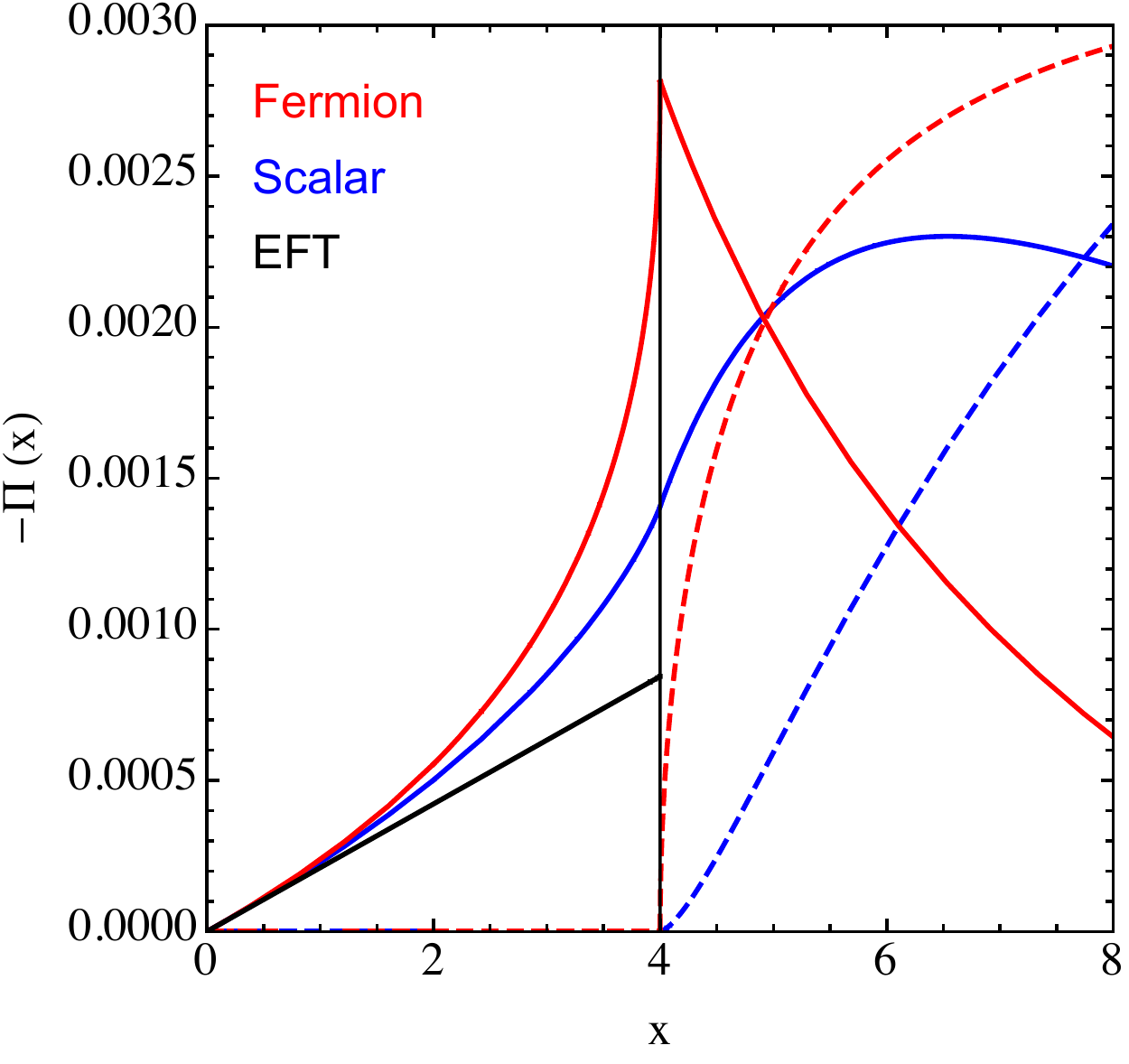}
\caption{\it Kinematical dependence of the form factor for fermions (red) and scalars (blue) running in the loop, 
and in the EFT limit (black).  
Full and dashed lines denote respectively real and imaginary part of the form factor. 
}\label{Plot:FFs}
\end{center}
\end{figure}

The contribution of $\chi$ to the EW gauge boson 
propagators is purely transversal and the $\bar{\text{MS}}$ renormalized form factors are 
(respectively for the case of a scalar and a fermion running in the loop): 
\begin{align}
\label{PixcalcS}
\Pi_S(x) &= 
- \frac{3 x \log \left(\frac{\mu ^2}{m_{\chi}^2}\right)+8 (x-3) +3 x
   \left(\frac{x-4}{x}\right)^{3/2} \log \left(\frac{1}{2}
   \left(\left(\sqrt{\frac{x-4}{x}}-1\right) x+2\right)\right)
   }{144 \pi ^2 x} \, ,  \\
\label{PixcalcF}
\Pi_F(x) &= 
-\frac{3 x \log \left(\frac{\mu ^2}{m_{\chi}^2}\right)+ 12 +5 x+3 \sqrt{\frac{x-4}{x}}
   (x+2) \log \left(\frac{1}{2} \left(\left(\sqrt{\frac{x-4}{x}}-1\right)
   x+2\right)\right)}{288 \pi ^2 x} \, .  
\end{align}
Here $x=q^2/m_{\chi}^2$, where $q$ is the external momentum of the 
gauge boson propagator and $\mu$ is the renormalization scale. 
A useful choice is $\mu = m_\chi$, which ensures that the form factors $\Pi_{S,F}$ vanish
for $x=0$. This choice is henceforth assumed. The behavior of the $\Pi_{S,F}$ form factors is shown in \fig{Plot:FFs}.

In the EFT limit, $x \ll 1$, the expanded form factor is $\Pi (x) \simeq  -x/(480 \pi ^2)$, 
both for scalar and fermions. 
Since $\Pi(0) = 0$ there is no contribution to the oblique parameters 
$S$, $T$, $U$ \cite{Peskin:1991sw}, while $W$ and $Y$ \cite{Barbieri:2004qk}, 
which correspond to the Wilson coefficients of the dimension-$6$ operators
$- \frac{W}{4 m^2_W} \left( D_\rho W^a_{\mu\nu} \right)^2$ 
and $- \frac{Y}{4 m^2_W} \left( \partial_\rho B_{\mu\nu} \right)^2$,   
are given by
\beq
\label{defWY2}
W = \frac{g^2 C^{\rm eff}_{WW}}{960 \pi^2} \frac{m^2_W}{m_\chi^2} \, , \qquad 
Y = \frac{g'^2 C^{\rm eff}_{BB}}{960 \pi^2} \frac{m^2_W}{m_\chi^2} \, .  
\eeq
For $x \gtrsim 1$ the EFT breaks down and hence 
the full momentum dependence of the form factor must be taken into account.
For $x \geq 4$ the momentum is above the pair-production threshold and the form factors develop an imaginary part (cf.~\fig{Plot:FFs}).

It is interesting to notice that for $1 \lesssim x \lesssim 4$ the full form factors are (significantly) larger than the
EFT approximation (compare the blue and red lines with the black line in \fig{Plot:FFs}).
This means that indirect searches for multiplets with a mass close to the pair production threshold tend
to be significantly more sensitive than what the EFT approximation would suggest.

\subsection{Modification of the SM amplitude}

In order to derive the radiative corrections to the neutral and charged current $2 \to 2$ fermion processes, 
we project \eq{LEWIMPs} onto the gauge boson mass eigenstates $\gamma,Z,W$
\beq 
\label{LEWIMPsBP}
\mathcal{L}_{\rm eff} = 
\mathcal{L}_{\rm SM} + 
\sum_{V,V'=\gamma,Z} 
\frac{d_{VV'}}{4} V_{\mu\nu} \Pi(-\partial^2/m_\chi^2) V'^{\mu\nu}
+ \frac{d_{WW}}{2} W^+_{\mu\nu} \Pi(-\partial^2/m_\chi^2) W^{-\mu\nu} \, ,
\eeq 
where 
$d_{\gamma\gamma} = (e^2/2) (C^{\rm eff}_{WW} + C^{\rm eff}_{BB})$,
$d_{ZZ} = (g_Z^2/2) (\cos^4\theta_W C^{\rm eff}_{WW} + \sin^4\theta_W C^{\rm eff}_{BB})$,  
$d_{\gamma Z} = d_{Z \gamma} = (e g_Z/2) (\cos^2\theta_W C^{\rm eff}_{WW} - \sin^2\theta_W C^{\rm eff}_{BB})$, 
$d_{WW} = (g^2/2) C^{\rm eff}_{WW}$, and we used the definitions $e=g \sin\theta_W$, $g_Z=g / \cos\theta_W$ 
and $\tan\theta_W = g'/g$. 
The modified EW gauge boson propagators for neutral and charged currents read respectively 
(keeping only the transverse part which is affected by new physics)
\begin{align}
\label{DeltaVV}
\Delta^{VV'}_{\mu\nu} &= i 
\left[ 
\begin{array}{cc}
(1-d_{\gamma\gamma}\Pi(x)) q^2 & - d_{\gamma Z} \Pi(x) q^2 \\
- d_{Z \gamma} \Pi(x) q^2 & (1-d_{ZZ}\Pi(x)) q^2 - m^2_Z
\end{array}
\right]^{-1}
\left(-g_{\mu\nu}+\frac{q_\mu q_\nu}{q^2}\right)
\, , \\
\label{DeltaWW}
\Delta^{WW}_{\mu\nu} &= \frac{i}{q^2 - m^2_W - d_{WW} q^2  \Pi(x)} 
\left(-g_{\mu\nu}+\frac{q_\mu q_\nu}{q^2}\right)
\, . 
\end{align}
Given the SM amplitude for a $2 \to 2$ fermion process, the effects of the new particle $\chi$ 
can be systematically accounted for by substituting the tree-level EW gauge boson propagators 
with the modified ones in \eqs{DeltaVV}{DeltaWW}. The leading correction to the SM 
cross-section comes from the interference with the SM amplitude, therefore it is due to 
real part of the form factor.

Note, also, that the contribution of the coefficient $C^{\rm eff}_{WW}$ 
typically gives the strongest constraint 
(this is for instance the case for the EW states introduced in \sect{physcaseew}). 
The reason depends in part on the hierarchy of the gauge couplings $g^2/g'^2 \sim 3$ 
and, more in general, on the fact that in the large $n$ and $y$ limit 
the effective coefficients scale like $C^{\rm eff}_{WW} \sim n^3$ and 
$C^{\rm eff}_{BB} \sim n y^2$. So, unless the hypercharge is exotically large 
(or both $n$ and $y$ are exotically large) the contribution of $C^{\rm eff}_{BB}$ is subleading. 
Moreover, one expects that for some large values of $n$ or $y$ perturbativity breaks down. 
This issue is analysed in \app{App:pert}, 
where we find that $n$ up to 9 can still be considered to be 
in the perturbative domain. 

Although we will present the mass exclusions as a function of 
the dimensionality $n$ of an irreducible SU(2)$_L$ representation, 
it is possible to recast our results for a generic SU(2)$_L$ reducible representation 
by properly rescaling the coefficient $C^{\rm eff}_{WW}$ (cf.~\eq{eq:def_C}). 
E.g.~$N$ copies of fundamentals ($n=2$) with degenerate mass 
would effectively correspond to a single representation 
with $n_{\star}$ obtained by solving $n_{\star}^3 - n_{\star} = N (2^3 - 2)$.

\section{Prospects at the HL-LHC and future hadron colliders}\label{sec:HL-LHC}

In this section we report the numerical results of our analysis in the context of hadron colliders. 
First of all we focus on the HL-LHC, then we consider possible future high-energy hadron
colliders, in particular the HE-LHC and FCC-100.

\subsection{Description of the analysis}

We perform a simple analysis based on a cut-and-count strategy. In the case of the neutral $\ell^+ \ell^-$
process we exploit the distribution in the invariant mass of the lepton pair, whereas in the case of the charged process
$\ell \nu$ we consider the distribution in the transverse mass of the event (one could equivalently use the
$p_T$ distribution of the charged lepton).

For simplicity we do not take into account the angular and rapidity distributions, which could be accessed in the
$\ell^+ \ell^-$ process. We expect this information not to play a relevant role in our analysis. In fact, since the leading
effects of EW multiplets at hadron colliders are driven by contributions to $C_{WW}^{\rm eff}$,
the angular and rapidity distributions are only marginally distorted. The contributions coming from $C_{BB}^{\rm eff}$,
which could be more easily distinguished through an angular analysis, are instead very suppressed and can
typically be neglected.

For both invariant mass and transverse mass distributions we use a binned log-likelihood analysis. The size of the bins
is chosen to be $15\%$ of their lower boundary and we only considered events above the $200\;$GeV threshold in
order to avoid large effects from real $Z$ or $W$ production. The boundaries of the bins are thus given by
$(200\;{\rm GeV})\times 1.15^n$. For the HL-LHC we include in the analysis bins up to $\sim 2\;$TeV, while for the HE-LHC
and FCC-hh we stop at $\sim 3.5\;$TeV and $\sim 20\;$TeV respectively.

The binned cross section at LO and NNLO QCD~\cite{Anastasiou:2003yy, Anastasiou:2003ds, Melnikov:2006di, Catani:2009sm, Catani:2010en} has been evaluated through the code {\tt FEWZ 2.0}~\cite{Gavin:2010az}, using the {\tt NNPDF30} PDFs~\cite{Ball:2014uwa} 
with $\alpha_s(M_Z)=0.118$.

\subsection{Results: HL-LHC}

We now summarize the results we obtained for the HL-LHC. In order to properly assess the sensitivity of our analysis
it is important to make realistic assumptions about the theoretical and experimental systematic uncertainties.
As benchmark targets we include in our analysis an uncorrelated and
a fully correlated systematical error both at the level of $2\%$ in the $\ell^+\ell^-$ channel and $5\%$ in the
$\ell\nu$ channel. These assumptions were found to be realistic in Ref.~\cite{Farina:2016rws}. We also separately include
the uncertainty in the PDFs. For simplicity we sum in quadrature the various errors.

\begin{figure}[t]
\includegraphics[width=0.48 \textwidth]{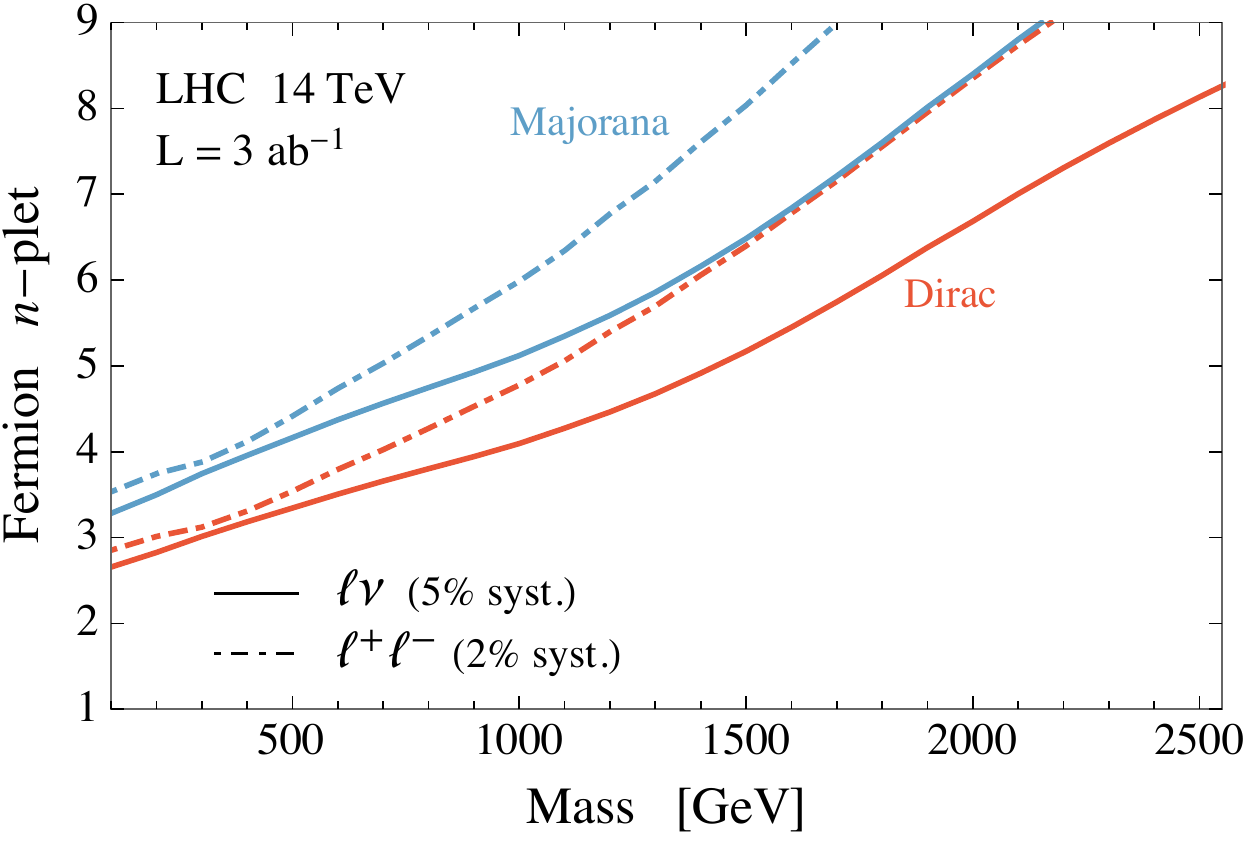}
\hfill
\includegraphics[width=0.485 \textwidth]{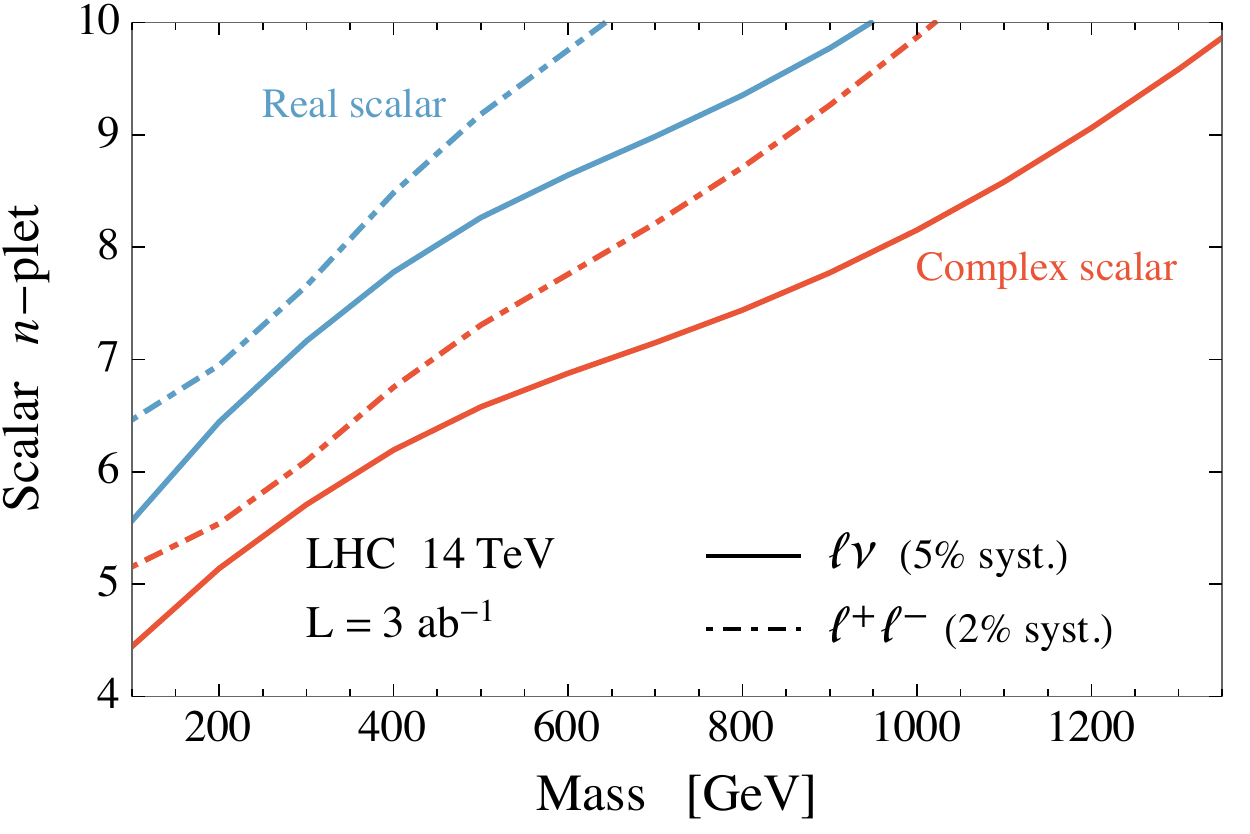}
\caption{\it Expected $95\%$ CL exclusion limits at the HL-LHC. The left and right panels show the bounds on fermion and scalar multiplets
respectively. The vertical axis reports the effective $n$ of the multiplet, while the horizontal axis gives the mass of
the states in the multiplet, which are assumed to be (almost) degenerate.
The solid and dot-dashed lines correspond to the bounds from the $\ell \nu$ and $\ell^+\ell^-$ channels respectively.
The blue (red) lines give the bounds for Majorana (Dirac) fermions on the left panel and for real (complex) scalars
in the right panel.}\label{fig:results_HL-LHC}
\end{figure}

The projection for the $95\%$ CL exclusion bounds for fermion and scalar multiplets are shown in
Fig.~\ref{fig:results_HL-LHC}. The plots show the exclusions on the ``effective size'' of the multiplet
as a function of the mass of the multiplet. For all multiplets we are considering, the
main corrections to the SM cross sections come from deformations of the ${\rm SU}(2)_L$ gauge propagator
(namely contributions to the $C_{WW}^{\rm eff}$ coefficient), while the effects of the hypercharge coupling are
practically negligible. The effective size is thus defined by converting the bound
on $C_{WW}^{\rm eff}$ into a bound on $n$ (considered now as a real number) through \eq{eq:def_C}.

The plots report the bounds from both the $\ell\nu$ channel (solid lines) and the
$\ell^+\ell^-$ channel (dot-dashed lines). One can see that the charged channel gives always the best sensitivity.
The difference is more noticeable for large multiplets ($n \gtrsim 4$), while it is milder for small multiplets
($n \sim 3$). The stronger sensitivity in the $\ell\nu$ channel is due to a combination of factors. First of all the
size of the deviations in this channel are slightly larger than in the $\ell^+\ell^-$ one. Moreover the cross section
in the charged channel is larger, thus providing a significant improvement on the statistics, which
helps especially for high masses, where the statistical uncertainty dominates over the systematic ones.

It is interesting to notice that, due to the sizable systematic uncertainty, hadron colliders cannot test small
${\rm SU}(2)_L$ multiplets even for small masses. This is due to the fact that the size of the fractional deviation
with respect to the SM cross section is fully determined by the multiplet size. For small multiplets these effects
are smaller than the systematic uncertainties and therefore not detectable.
We will see in the following (\sect{sec:syst_unc}) how a change in the systematic uncertainties affects these results.

The expected exclusion bounds for several multiplets which could provide a DM candidate are reported in \Table{tableMDM}.
One can see that the HL-LHC does not allow to test any of the multiplets for mass values which saturate the
DM relic abundance.

\subsubsection{LO vs.~NNLO}

\begin{figure}[t]
\centering
\includegraphics[width=0.5 \textwidth]{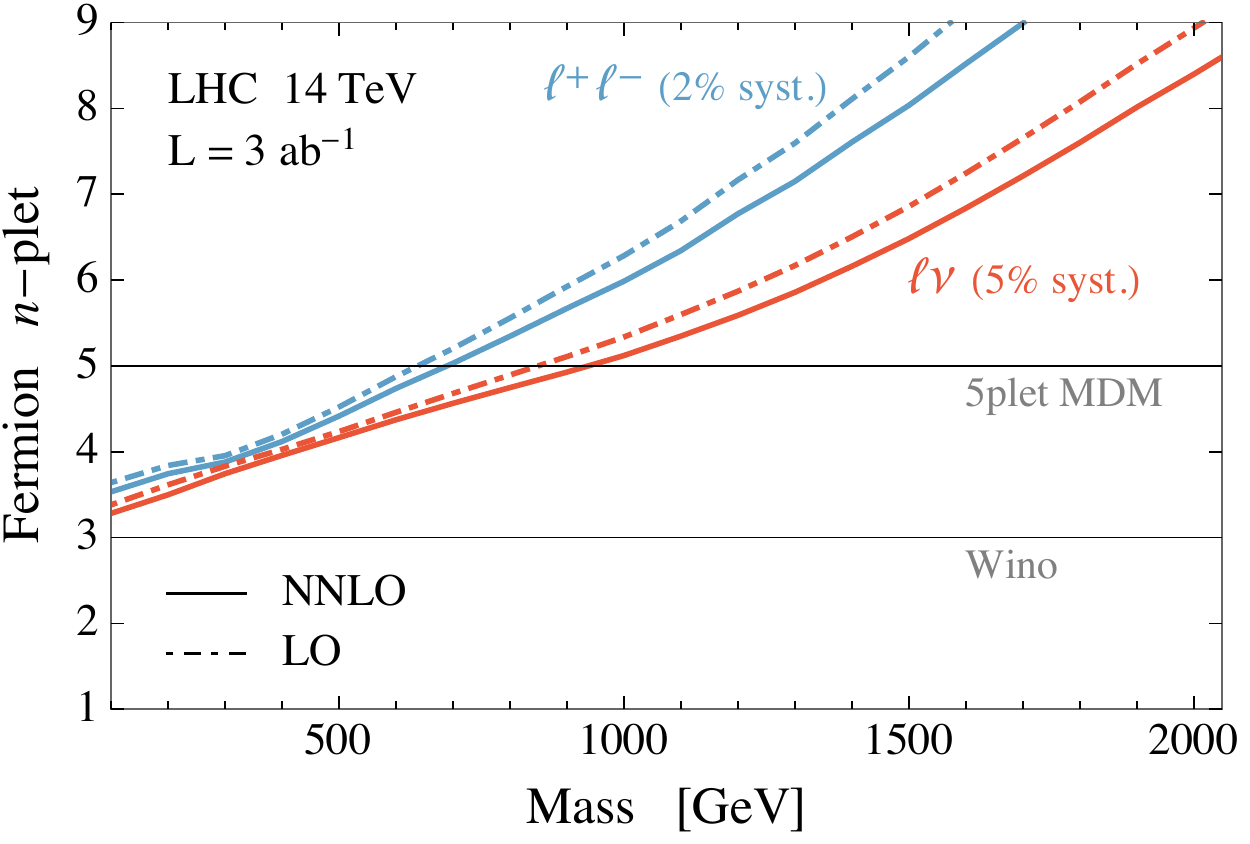}
\caption{\it Comparison of the $95\%$ CL exclusion bounds on Majorana fermion multiplets at the HL-LHC
obtained by using the LO (dot-dashed lines) and NNLO (solid lines) distributions.
The blue lines refer to the $\ell^+\ell^-$ process, while the red ones correspond to the $\ell \nu$ process.}\label{fig:LOvsNNLO}
\end{figure}

It is interesting to study the impact of higher-order corrections to the kinematic distributions on the
exclusion bounds. For this purpose we show in \fig{fig:LOvsNNLO} how the bounds change if we use LO
or NNLO distributions to derive them.

The figure shows that the impact of higher-order corrections is relatively mild, giving an increase in the mass bound
of order $5\% - 10\%$. This is due to the fact that the higher-order corrections give a mild enhancement
(of order $20\%$) of the cross section. Notice that most of the enhancement is just given by the NLO QCD contributions,
while the impact of the NNLO corrections on the extraction of the bounds
are practically negligible. For this reason in \fig{fig:LOvsNNLO} we did not report
the NLO lines, which almost exactly overlap with the NNLO ones.

\subsubsection{Impact of systematic uncertainties}\label{sec:syst_unc}

\begin{figure}[t]
\includegraphics[width=0.48 \textwidth]{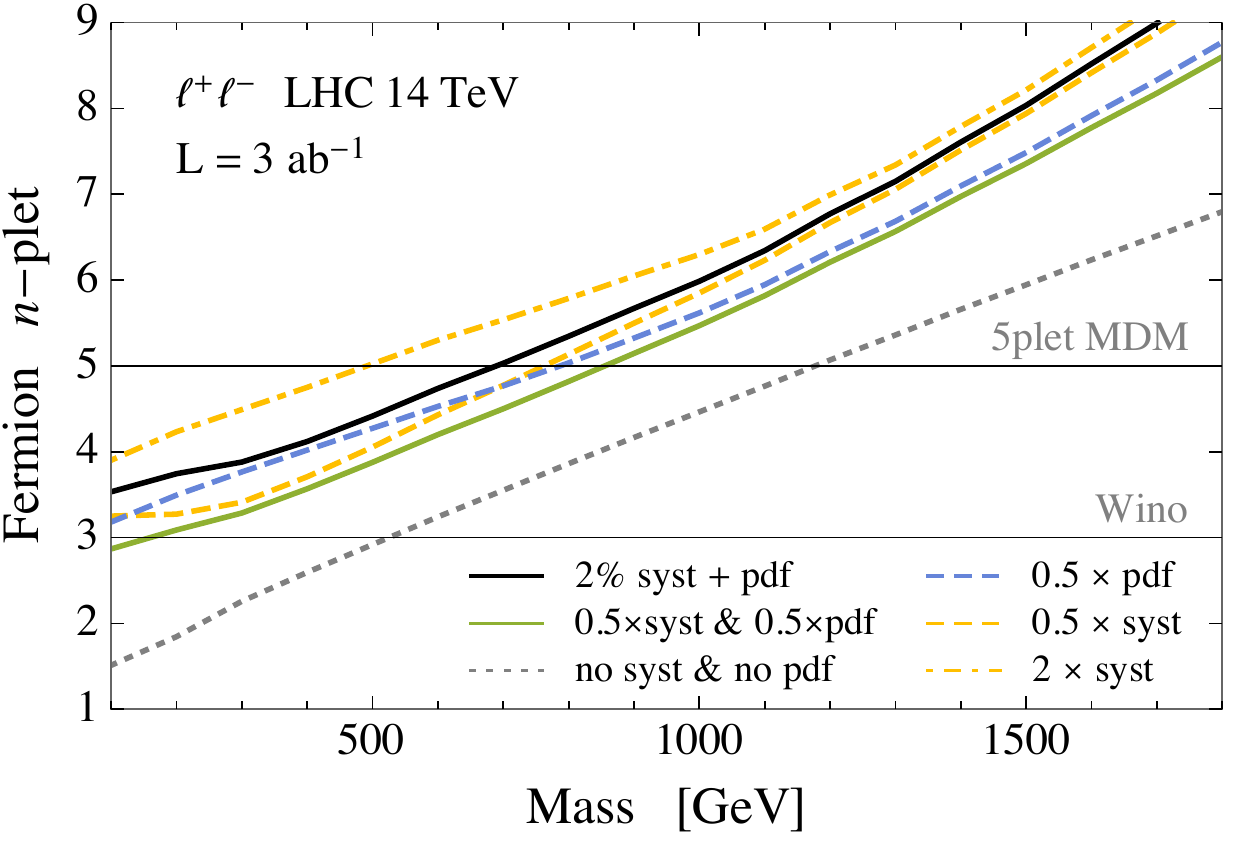}
\hfill
\includegraphics[width=0.48 \textwidth]{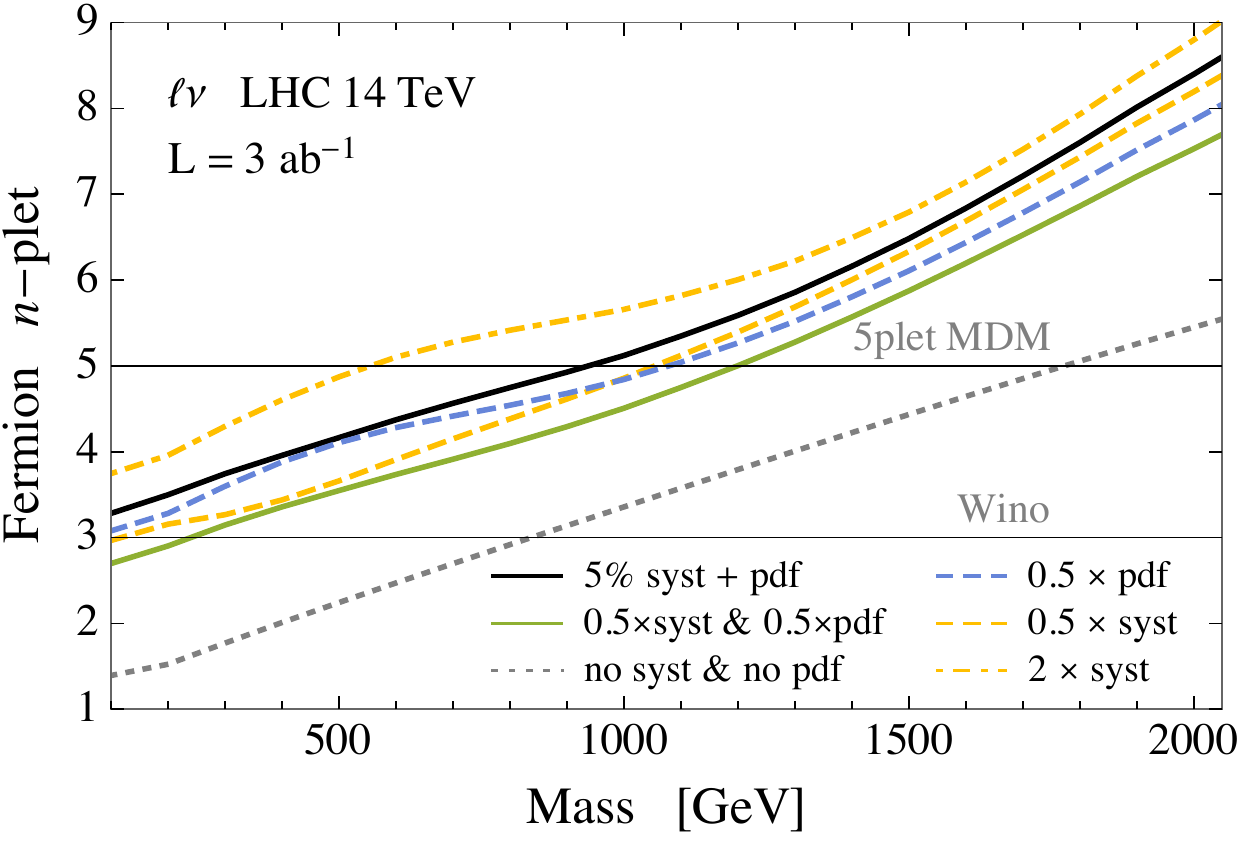}
\caption{\it Dependence of the $95\%$ CL exclusion bounds on Majorana fermion multiplets at the HL-LHC
on the PDF uncertainties and on the additional systematic errors. The left and right panels show the
bounds from the $\ell^+\ell^-$ and $\ell\nu$ channels respectively. The solid black lines give the bounds for
our benchmark systematic errors. The dashed (dot-dashed) yellow lines show how the bound changes if the
additional systematic uncertainties are halved (doubled). The dashed blue lines show how the bounds are modified
by halving the PDF uncertainty. The solid green lines give the bound with all systematic 
uncertainties
halved. Finally the dotted grey lines correspond to the bounds with no systematic and PDF errors.
}\label{fig:syst_unc}
\end{figure}

A second aspect we want to discuss is the dependence of our results on the systematic uncertainties. Focusing
again on the Majorana fermion case, we show in \fig{fig:syst_unc} how the exclusion bounds change
by varying the different sources of systematic uncertainty, namely the PDF uncertainty and
the additional correlated and uncorrelated errors.

One can clearly see that the impact of the PDF uncertainty is dominating for large multiplet masses, roughly above
$1\;$TeV. In this region the PDF uncertainty becomes much larger than the other sources of systematic error.
The impact of halving the PDF uncertainty provides an increase in the bounds by roughly $10\% - 15\%$ for high
masses.

On the other hand, the additional systematic uncertainties dominate for low masses. 
The dot-dashed yellow lines show
that a factor-of-two increase of these uncertainties can significantly degrade the bounds for masses $\lesssim 1\;$TeV, while 
a reduction by a factor of two allows to access significantly lower $n$ for small multiplet
masses or, equivalently, could allow for even a $\sim 100\%$ improvement in the bounds at fixed $n$.

One can see that an overall reduction by $50\%$ of all systematic uncertainties can give a significant improvement in the
exclusion reach. In particular it allows us to probe the $n=3$ case, which can be interpreted as a
mass-degenerate wino multiplet. This class of states is not accessible in the benchmark systematic
error scenario we were considering for our main results.

Finally, mainly for illustrative purposes, we also report the bounds obtained by neglecting all systematic uncertainties
(dashed grey lines). In this case, thanks to the very high statistics in the low-mass bins, even an ${\rm SU}(2)_L$
triplet\footnote{Notice that a degenerate Higgsino multiplet would correspond (neglecting the 
subleading effect from the hypercharge) 
to an effective Majorana fermion with $n=2.4$ in \fig{fig:syst_unc}.} 
could be tested up to a mass $\sim 800\;$GeV.

\subsection{Results: Future hadron colliders}

We now present the expected exclusion limits at future high-energy hadron colliders, in particular HE-LHC and
FCC-100. We choose as benchmarks for the integrated luminosity $L = 10\;$ab$^{-1}$ for the HE-LHC and
$L = 20\;$ab$^{-1}$ for FCC-100. As for the HL-LHC case, we include in the analysis the present PDF errors,
as well as additional uncorrelated and fully correlated systematic uncertainties at the level of $2\%$ for the
$\ell^+\ell^-$ channel and $5\%$ for the $\ell\nu$ channel. Results for alternative luminosity benchmarks, as well
as for different choices of systematic uncertainties, are reported in \app{app:additional_results}.

\begin{figure}[t]
\includegraphics[width=0.48 \textwidth]{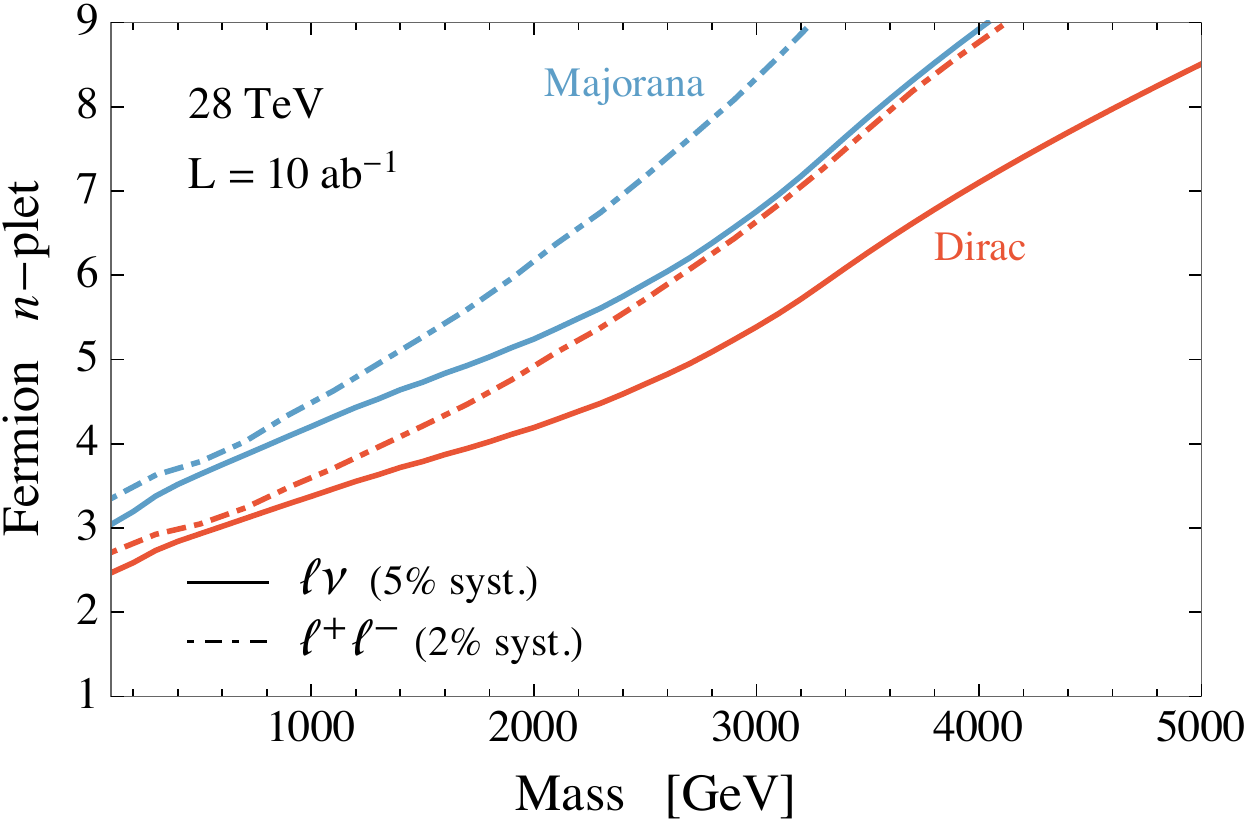}
\hfill
\includegraphics[width=0.48 \textwidth]{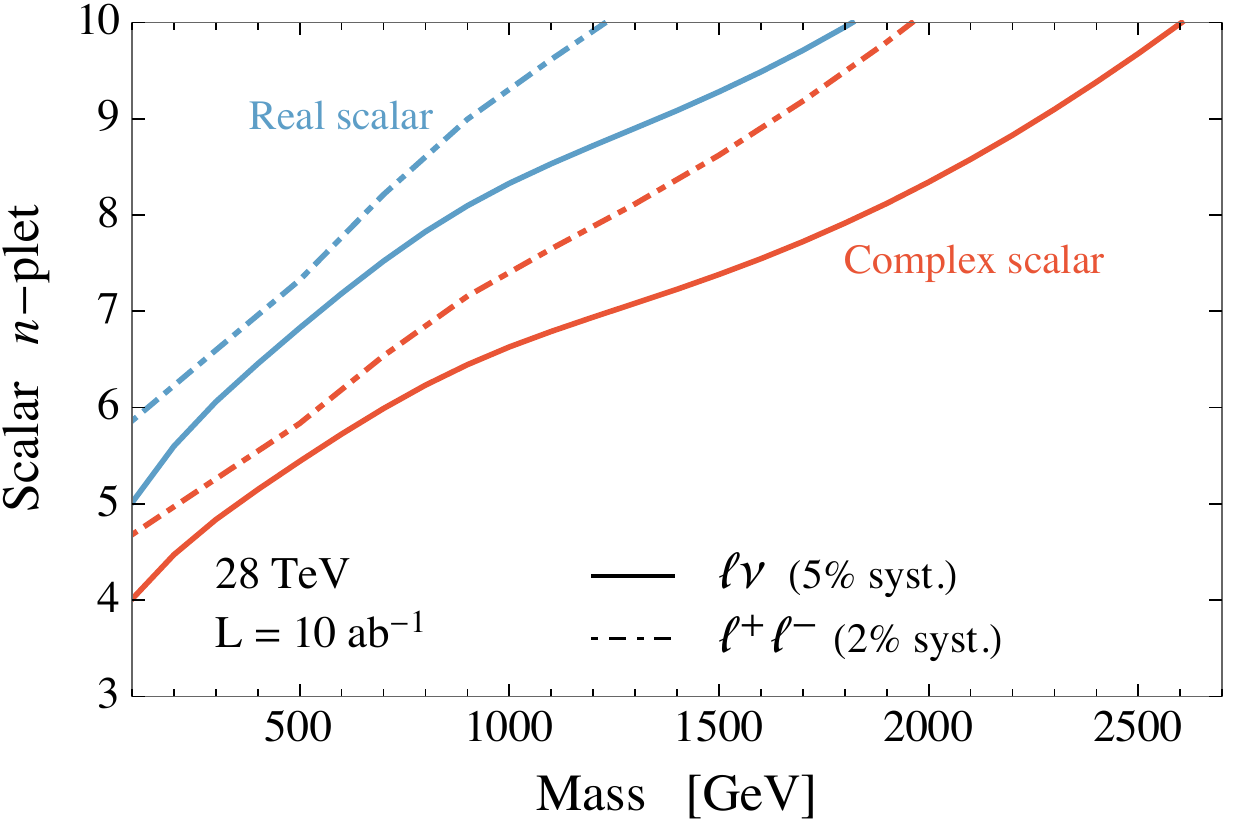}
\caption{\it Expected $95\%$ CL exclusion limits obtained for the HE-LHC.}\label{fig:results_HE-LHC}
\end{figure}

\begin{figure}[t]
\includegraphics[width=0.48 \textwidth]{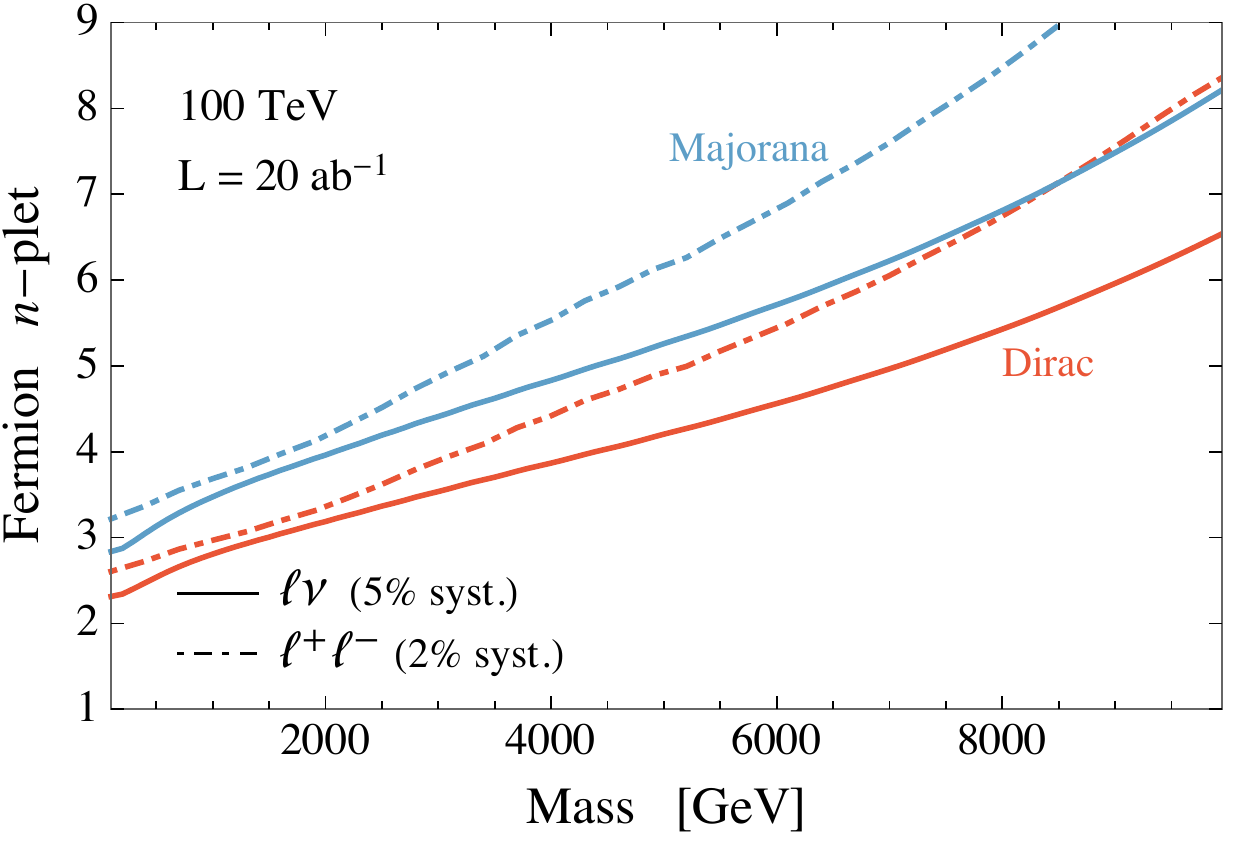}
\hfill
\includegraphics[width=0.49 \textwidth]{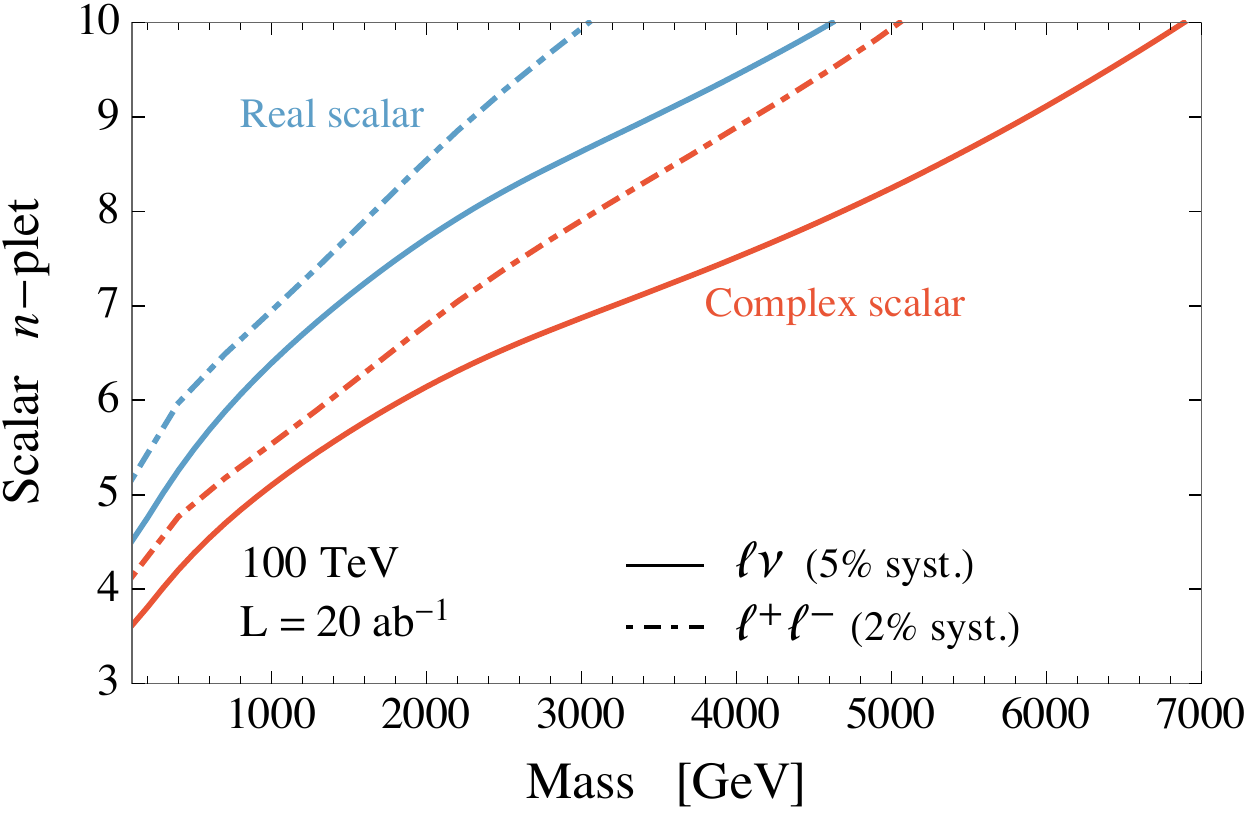}
\caption{\it Expected $95\%$ CL exclusion limits obtained for the FCC-100.}\label{fig:results_FCC}
\end{figure}

The expected bounds for the HE-LHC and FCC-100 benchmarks are shown in \fig{fig:results_HE-LHC}
and \fig{fig:results_FCC} respectively. Comparing with the HL-LHC results one can see that the HE-LHC
allows for an improvement of the mass bounds by roughly a factor $2$, whereas FCC-100 gives an improvement
by roughly a factor $5$. The advantage of a larger centre-of-mass energy translates mostly in shifting the
exclusion reach to higher masses, while the improvement in reaching smaller multiplet sizes is limited.
EW triplets, which were out of the HL-LHC reach, can be tested at high-energy hadron colliders
only for the case of Dirac fermions, whereas they remain outside the reach for the case of Majorana fermions
and for multiplets of scalars.

The expected exclusion bounds for multiplets which could provide a DM candidate are reported in \Table{tableMDM}.
The HE-LHC reach is still far from probing realistic DM masses. On the other hand FCC-100 could successfully test
the MDM scenario with a Dirac fermion $5$-plet $(1,5,\epsilon)_{\rm DF}$ and is not far from probing
the triplet $(1,3,\epsilon)_{\rm DF}$ and $7$-plet $(1,7,\epsilon)_{\rm DF}$ cases.

\section{Prospects at future lepton colliders}
\label{ewptCLIC}

The main difference of lepton machines compared to hadron colliders is the fact that 
the momentum of the EW gauge boson propagators is fixed by the centre-of-mass 
energy $\sqrt{s}$ (i.e.~it is not smeared by the PDF). 
Although the most interesting region for lepton colliders is the 
one below the energy threshold for $\chi$ pair production, EW corrections to 
$2 \to 2$ SM fermion processes could also be used to probe the region above threshold, 
in a complementary way with respect to direct searches. 
Other advantages of $\ell^+\ell^-$ machines are the reduced systematics and 
(in the case of electrons) the possibility of playing with beam polarization, 
whose main role is that of increasing the production cross-section. 

For definiteness we are going to consider two scenarios: an $e^+e^-$ collider 
inspired by the CLIC design and a futuristic high-energy muon collider.
As benchmarks for CLIC we focus on centre-of-mass energies foreseen for
the second and third stages of the experiment, 
namely 1.5 TeV (CLIC-2) and 3 TeV (CLIC-3) \cite{Aicheler:1500095}. 
The final luminosities 
after 27 years of run are estimated to be 
$2.5\text{ ab}^{-1}$ for CLIC-2 and 
$5\text{ ab}^{-1}$ for CLIC-3 \cite{talkCLIC}. 
One can further benefit from possible electron and
positron polarization. The cross-section of a generically polarized $e^+ e^-$ beam 
in terms of the polarization fractions $P_{e^-}$ and $P_{e^+}$ is defined by  
\begin{align}
\sigma_{P_{e^-}P_{e^+}} =& \frac{1}{4} 
\left[
(1+P_{e^-})(1+P_{e^+}) \sigma_{RR} + (1-P_{e^-})(1-P_{e^+}) \sigma_{LL} \right. \nonumber \\
& \left. + (1+P_{e^-})(1-P_{e^+}) \sigma_{RL} + (1-P_{e^-})(1+P_{e^+}) \sigma_{LR}  
\right] \, , 
\end{align}
where $\sigma_{LR}$ stands for instance for the cross-section if the 
$e^-$-beam is completely left-handed polarized 
($P_{e^-}=-1$) and the $e^+$-beam is completely right-handed polarized ($P_{e^+}=+1$). 
For our analysis it turns out to be helpful 
to have negative $e^-$ polarization and positive $e^+$ polarization, 
since this configuration enhances the $e^+e^- \to f \bar f$ cross-section.  
The Higgs program at CLIC prefers $P_{e^-}=-80\%$, while for a measurement of top quark 
couplings $P_{e^-}=+80\%$ is preferable. It can be assumed that 4/5 of the time will be
devoted to the Higgs program so we will use as a benchmark $P_{e^-}=-80\%$,
with an effective luminosity of $L=2\text{ ab}^{-1}$ for CLIC-2 and 
$L=4\text{ ab}^{-1}$ for CLIC-3 \cite{talkCLIC}. 
We further rescale the latter luminosities by a 0.6 factor, in order to account for 
beam effects.\footnote{We thank Jorge De Blas for correspondence about this point.}
There is also the possibility of positron polarization at a lower level, 
although positron polarization is not part of the baseline CLIC design 
\cite{Aicheler:1500095}. 
To this end we will consider either $P_{e^+}=0\%$ 
or $P_{e^+}=30\%$ (the latter corresponding to the configuration employed in \Table{tableMDM}). 

The option of a muon collider was recently revived\footnote{For a 
recent phenomenological analysis pointing out the advantages of a muon collider 
see Ref.~\cite{Buttazzo:2018qqp}.} thanks to new proposals for muon sources. 
In particular, the proposal of a low emittance muon source from positron scattering on a target,
LEMMA \cite{Antonelli:2015nla, Collamati:2017jww, Boscolo:2018tlu}, 
allows to reach centre-of-mass energies above 10 TeV \cite{talkwulzer}. 
As a benchmark we choose a 14 TeV
collider option with a luminosity of $L=20\text{ ab}^{-1}$. For comparison we 
vary the luminosity by factors of $1/4,\; 1/2$ and 4, and we consider even a higher energy option
of $\sqrt{s}=30\text{ TeV}$ for various values of the luminosity.\footnote{We thank Andrea Wulzer 
for correspondence about possible luminosity benchmarks.}

Following Ref.~\cite{Harigaya:2015yaa} we perform a binned likelihood analysis on the differential cross 
section of the process 
$\ell^+\ell^- \to f \bar{f}$ with respect to the cosine of the scattering angle $\theta$. In particular, we divide the latter in $10$ uniform intervals  
for $\cos\theta \in [-0.95, 0.95]$. 
For the final states we assume the following detection efficiencies: 
$100\%$ for leptons, $80\%$ for $b$-jets and $50\%$ for $c$-jets.
For our numerical analysis we compute the cross sections at LO. Our bounds are however in good agreement with the ones
of Ref.~\cite{Harigaya:2015yaa}, which includes NLO corrections. This shows that NLO effects have only a small impact on
the results.

\subsection{Results: $e^+e^-$ collider}

\begin{figure}[ht]
\begin{center}
\includegraphics[width=0.49 \textwidth]{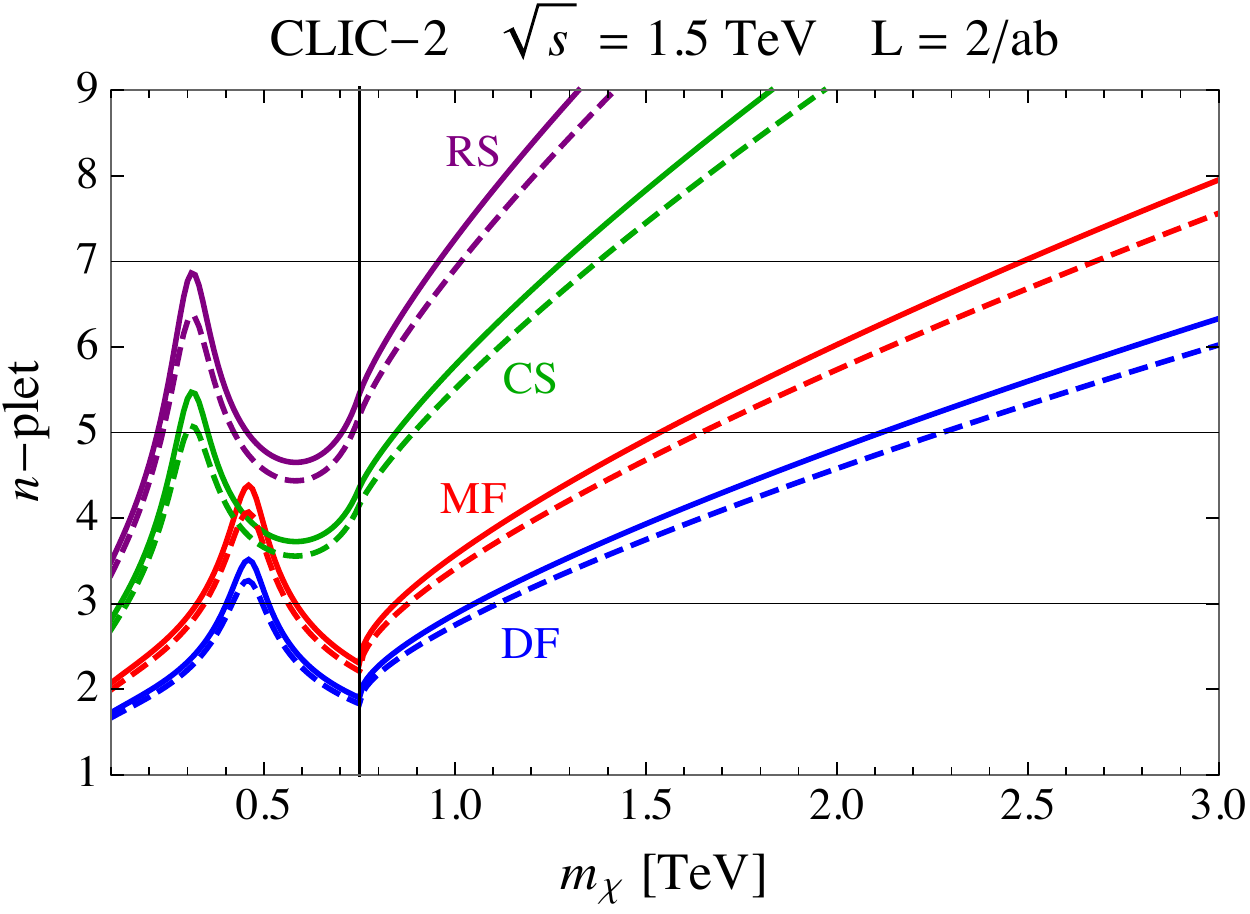} \
\includegraphics[width=0.49 \textwidth]{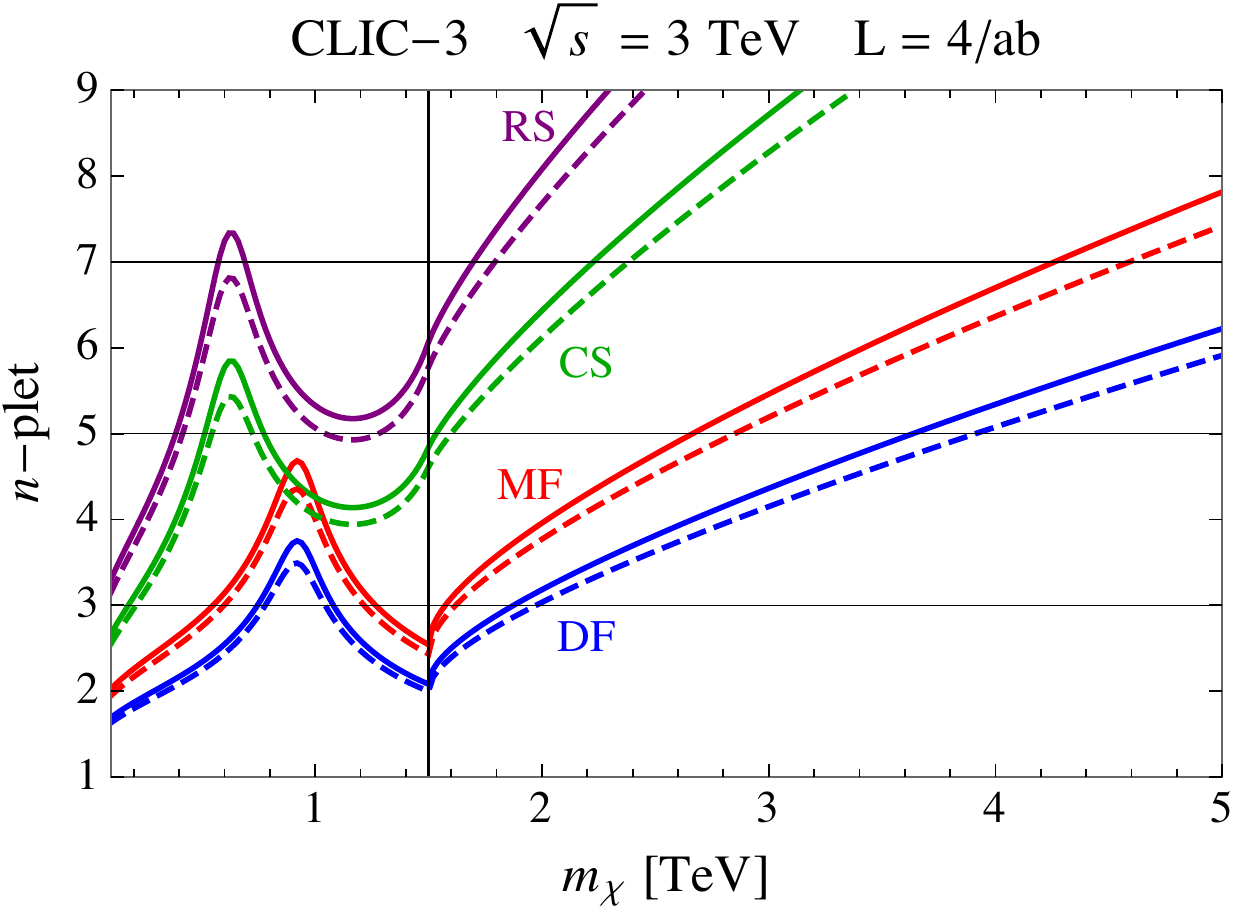}
\caption{\it Expected $95\%$ CL exclusion limits for CLIC-2 (left panel) and CLIC-3 (right panel), 
for different Lorentz representations  
and polarization fractions 
$(P_{e^-}, P_{e^+}) = (- 80 \%, 0)$ [full lines] and $(P_{e^-}, P_{e^+}) = (- 80 \%, + 30\%)$ [dashed lines]. 
}\label{Plot:Results_CLUC_comb2}
\end{center}
\end{figure}

The results are displayed in \fig{Plot:Results_CLUC_comb2} where we show the $95\%$ CL
exclusion limits in the plane $(m_\chi, n)$ for different Lorentz representations (RS, CS, MF, DF)
and for 
CLIC-2 ($\sqrt{s} = 1.5$ TeV, $L = 2 \ \text{ab}^{-1}$) and 
CLIC-3 ($\sqrt{s} = 3$ TeV, $L = 4 \ \text{ab}^{-1}$). 
To obtain these exclusions we have combined the $e/\mu/b/c$ channels 
assuming a systematic error of $0.3 \%$, we rescaled the luminosities by a 
0.6 factor due to beam effects and considered the polarization fractions $P_{e^-} = - 80 \%$ and 
$P_{e^+} = 0$ ($+30\%$).  

The vertical black line in both plots denotes the kinematical threshold for pair-production $\sqrt{s} / 2$. 
In the region below threshold (on the right side of the vertical black line) the bound on the mass grows with the 
dimensionality of the multiplet and eventually enters the EFT regime for $m_\chi \gg \sqrt{s}/2$ (cf.~\fig{Plot:FFs}). 

The bounds in the region above threshold (on the left side of the vertical black line) have a non-trivial 
behavior, showing a significant dip in sensitivity. This effect can be traced back to the fact that the real part of the form factor above threshold has an accidental zero
(e.g.~$x \equiv s / m_\chi^2 \simeq 11$ in the case of the self-energy correction due to fermions). 
For such value of $s/m_\chi^2$ the new-physics contributions are strongly suppressed since the interference with the SM
vanishes and the corrections only come from the square of the imaginary part of the $\Pi_{S,F}$ form factors.
Although a dip in sensitivity arises for certain values of $x$, one can still extract meaningful bounds 
also above the threshold for pair-production, which can be competitive with direct searches 
(cf.~also \sect{compDS}).
Due to the sensitivity dip, runs at different center of mass energy can be complementary in testing multiplets with
relatively low masses. For instance CLIC-3 can not test a $(1,3,0)_{\rm DF}$ multiplet for masses in the
range $[0.8, 1.0]\;$TeV. This mass range can however be fully tested at CLIC-2.

Regarding the sensitivity to DM candidates, we notice that CLIC-3 (together with CLIC-2)
would be able to cover the relevant parameter space of the $(1,3,\epsilon)_{\rm DF}$ multiplet
up to the thermal mass that saturates the DM relic abundance.

\subsection{Results: muon collider}

\begin{figure}[t]
\begin{center}
\includegraphics[width=0.44 \textwidth]{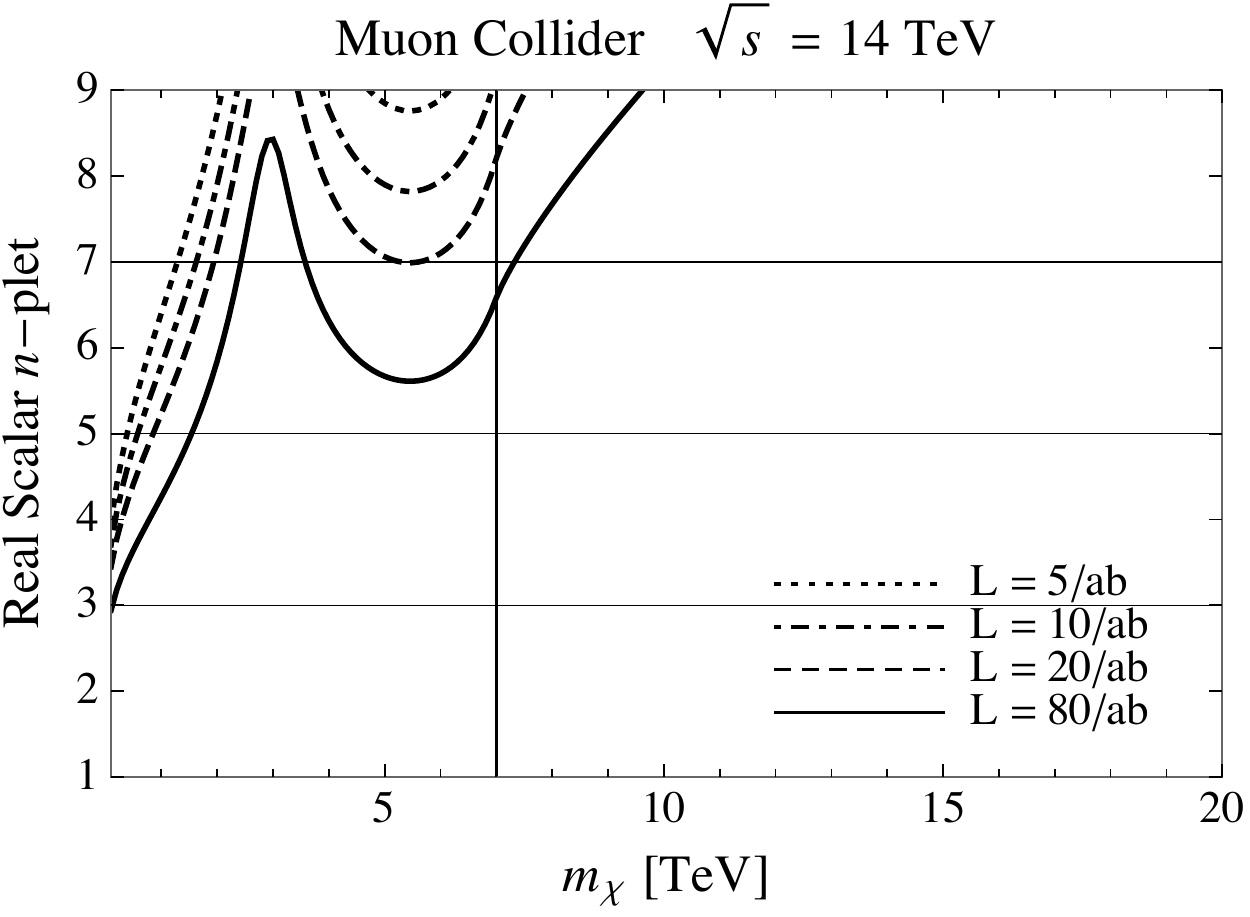}
\hspace{2em}
\includegraphics[width=0.44 \textwidth]{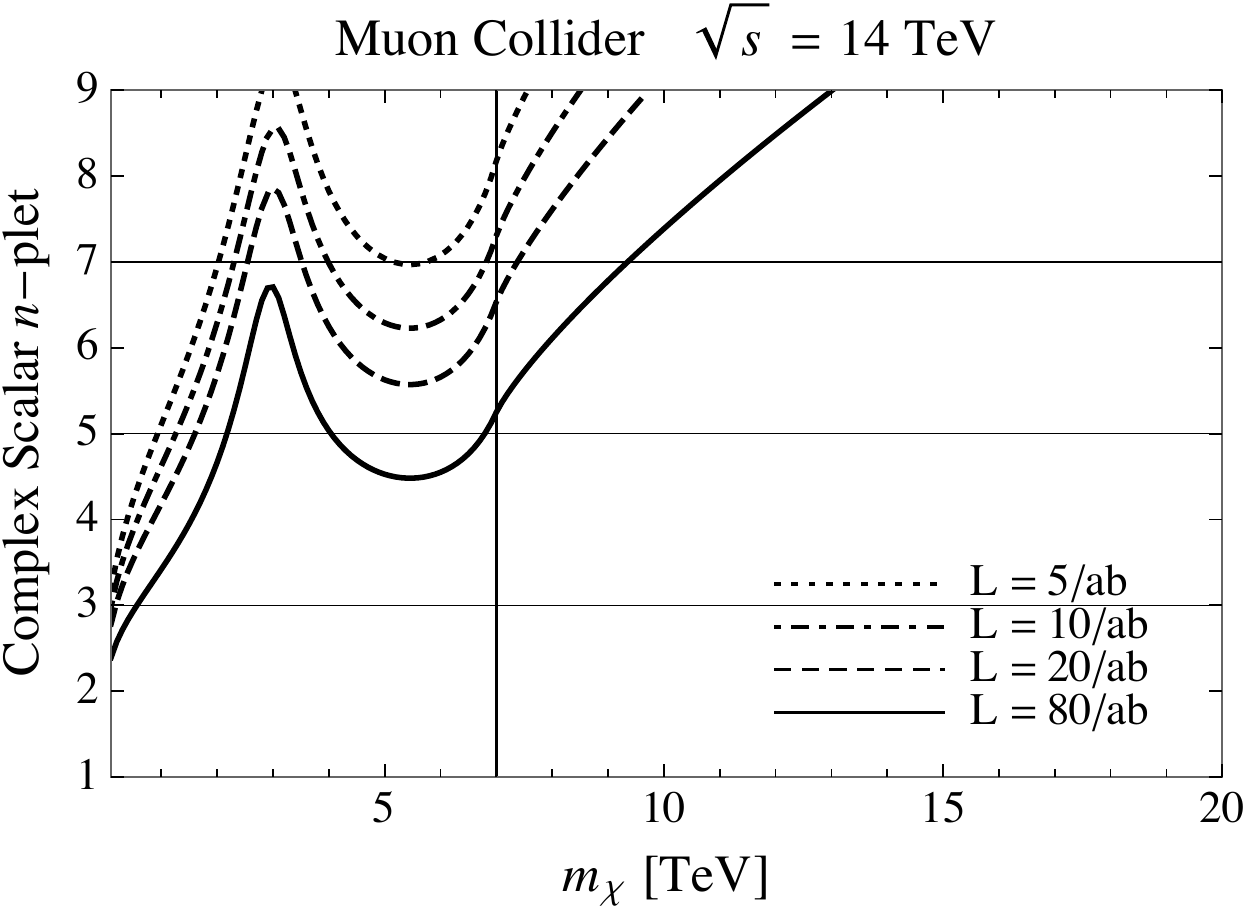} \\
\vspace{.1cm} 
\includegraphics[width=0.44 \textwidth]{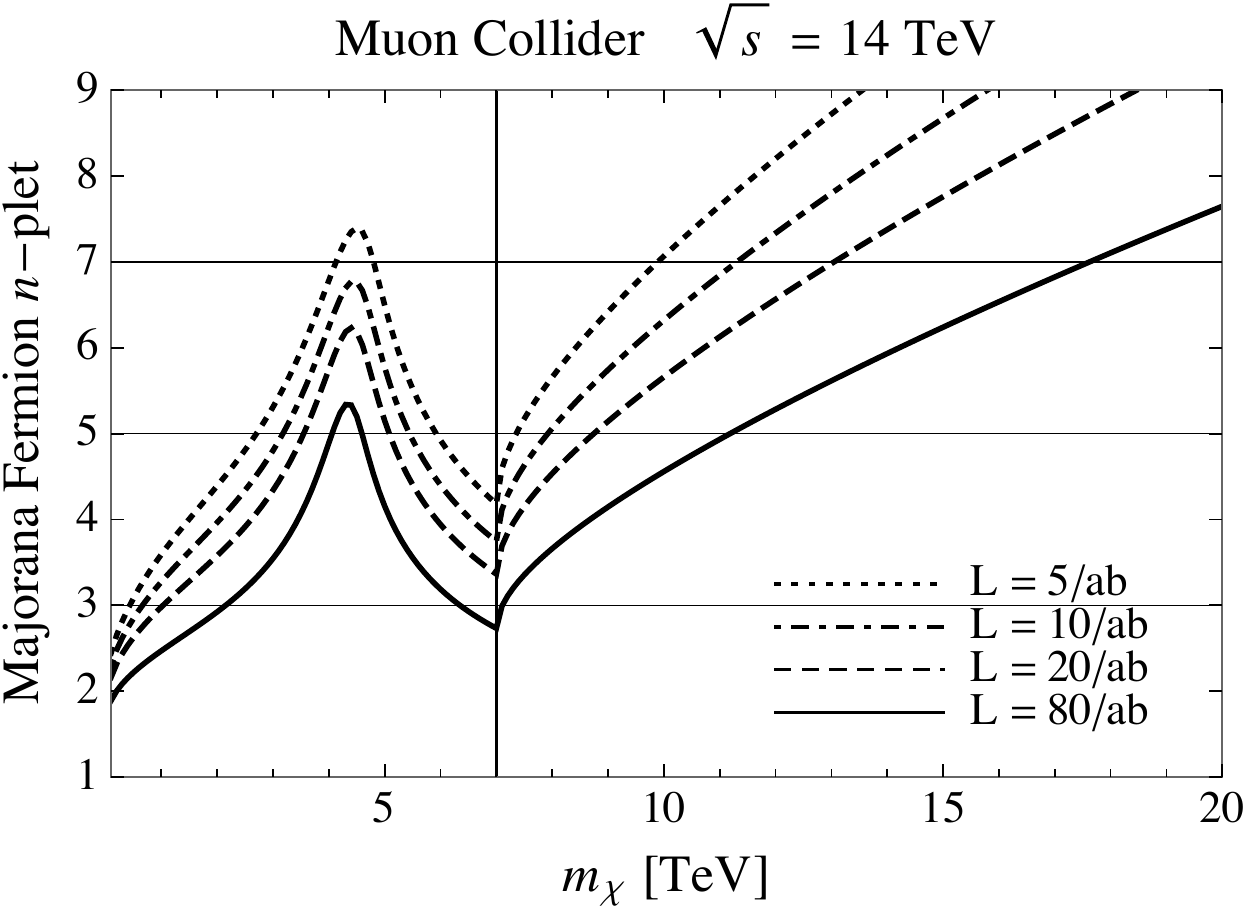}
\hspace{2em}
\includegraphics[width=0.44 \textwidth]{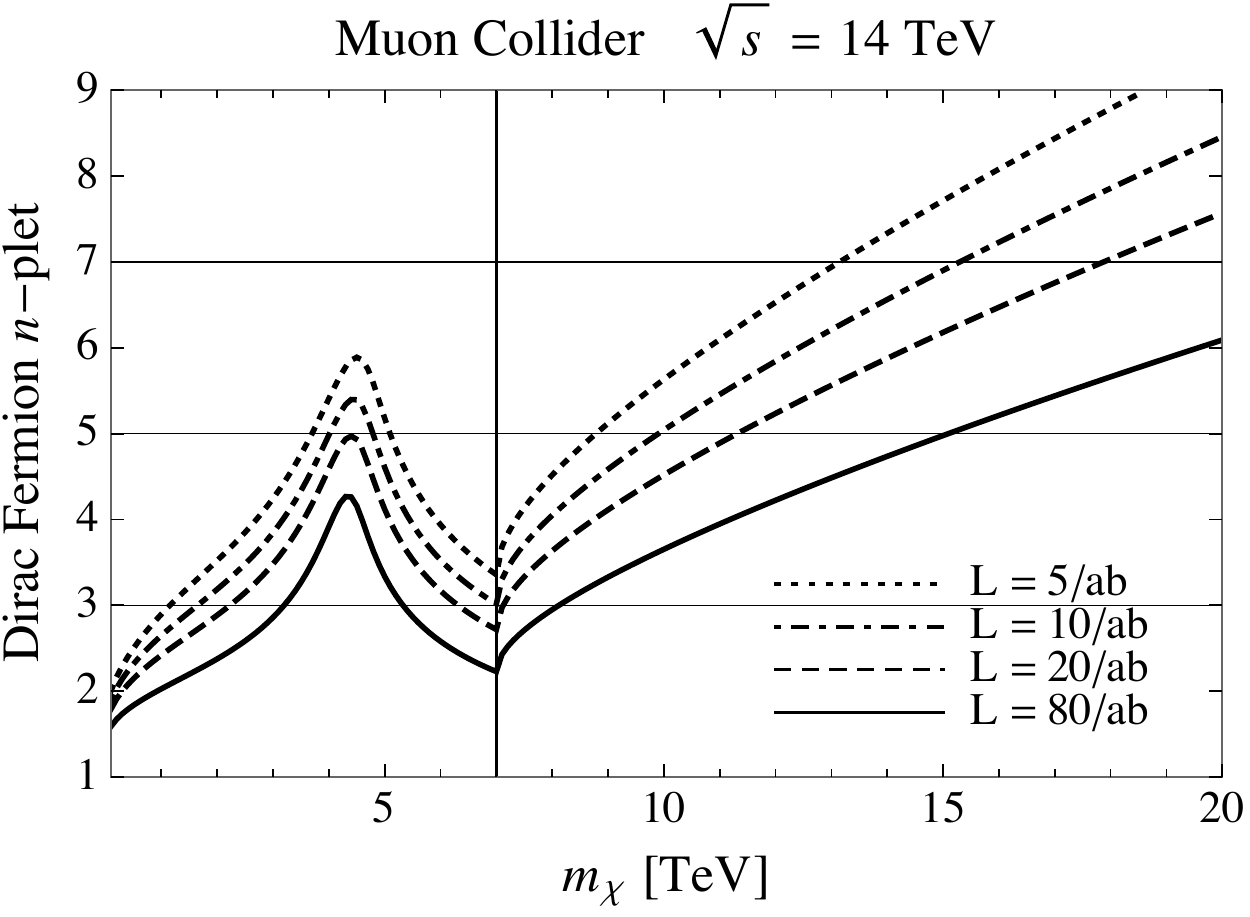} 
\vspace{-1em}
\caption{\it 
Expected $95\%$ CL exclusion limits for a 14 TeV 
muon collider.  
}\label{Plot:Results_muon_14TeV_differentL}
\vspace{1.25em}
\includegraphics[width=0.44 \textwidth]{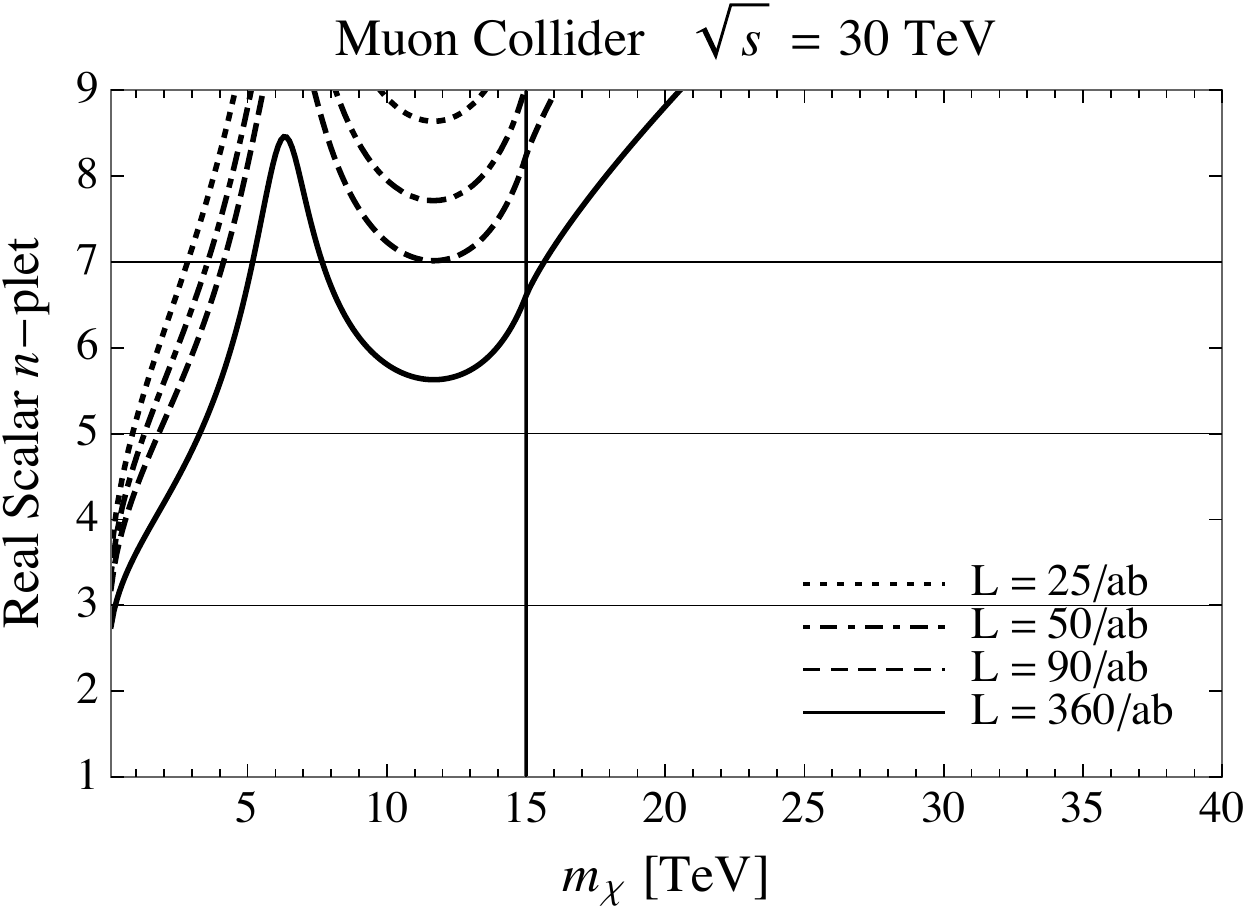}
\hspace{2em}
\includegraphics[width=0.44 \textwidth]{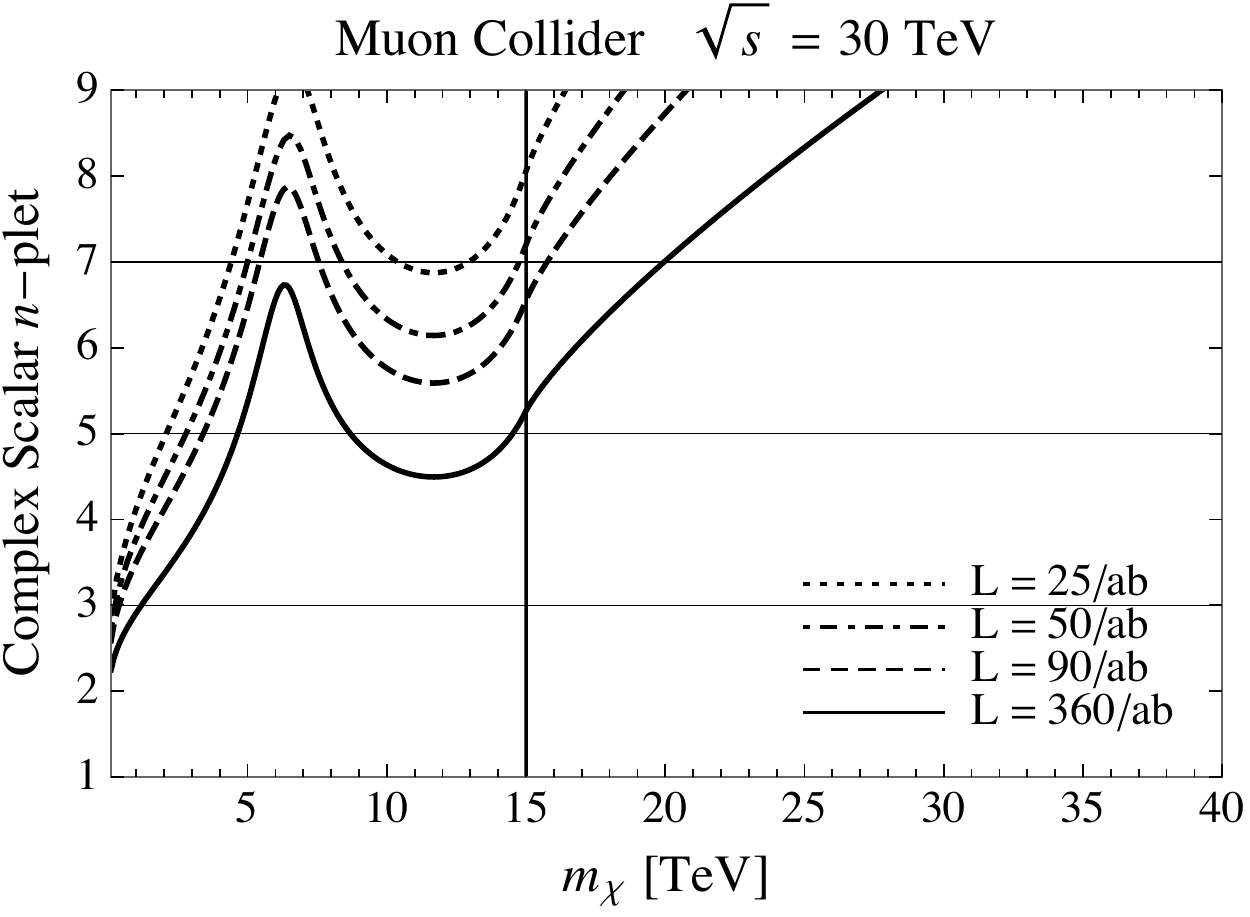} \\
\vspace{.1cm} 
\includegraphics[width=0.44 \textwidth]{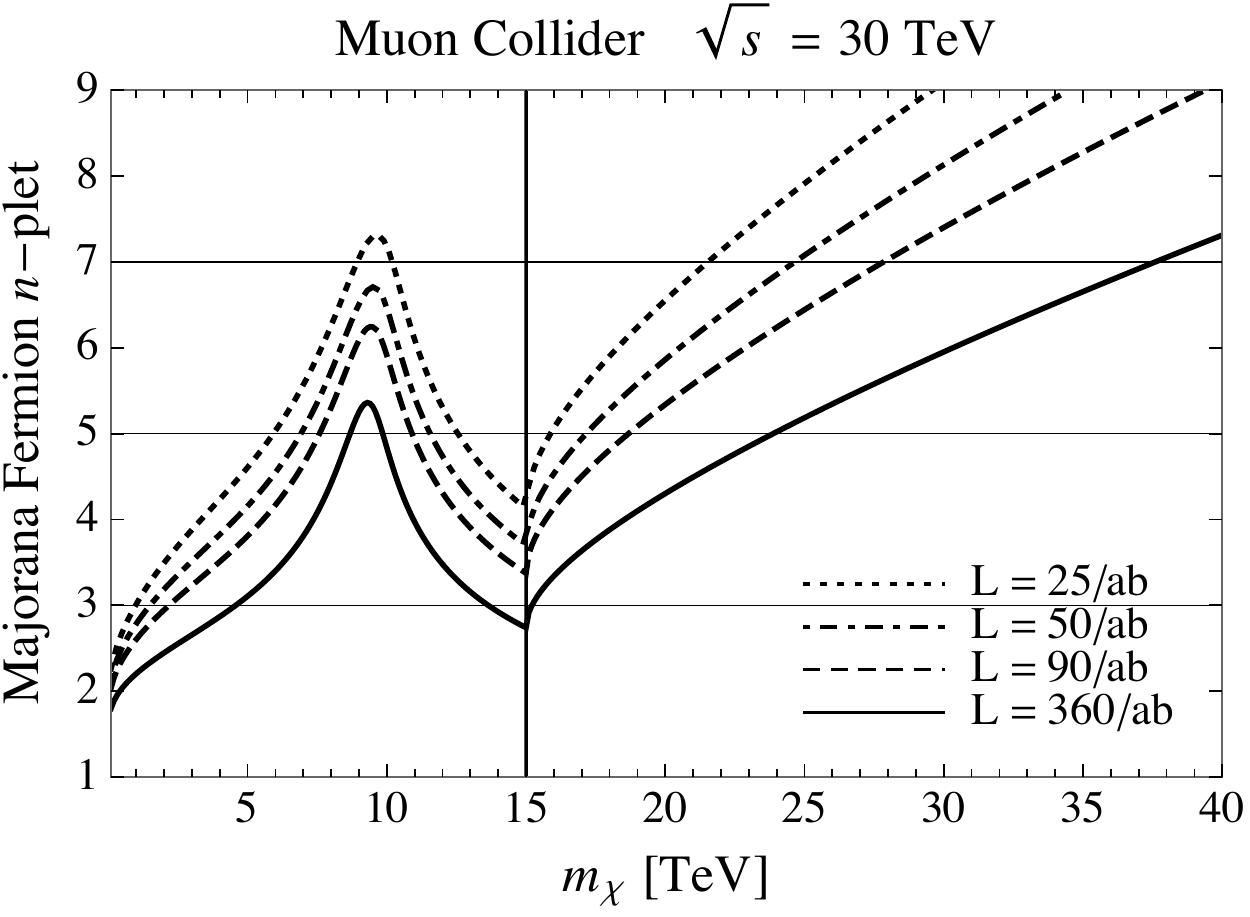}
\hspace{2em}
\includegraphics[width=0.44 \textwidth]{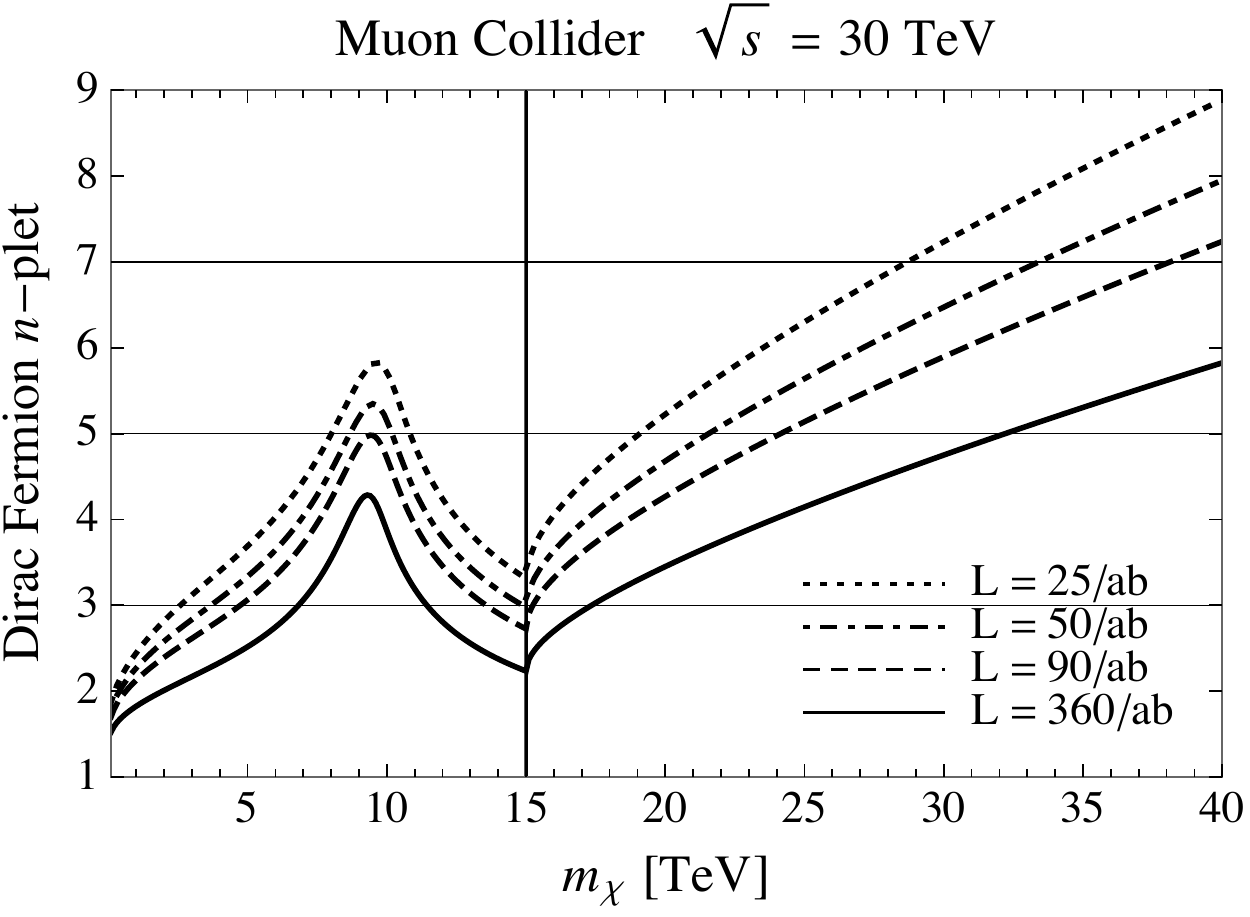}
\vspace{-1em}
\caption{\it 
Expected $95\%$ CL exclusion limits for a 30 TeV 
muon collider.  
}\label{Plot:Results_muon_30TeV_differentL}
\end{center}
\end{figure}

For the case of the muon collider we report the results in a slightly different way, 
emphasizing the role of luminosity. In particular, we report in 
\figs{Plot:Results_muon_14TeV_differentL}{Plot:Results_muon_30TeV_differentL}
the $95\%$ CL bounds for $\sqrt{s} = 14$ TeV and 30 TeV, and for various 
Lorentz representations and integrated luminosities (assuming a systematic error of $0.3 \%$ and no polarization).  

We notice that a high-energy muon collider can significantly extend the reach with respect to CLIC, 
provided enough integrated luminosity can be collected. 
Since the $\ell^+ \ell^- \to f \bar f$ cross section decreases quadratically with the energy,
a quadratic increase in the integrated luminosity is needed in order to allow for comparable sensitivity.
Smaller luminosities can instead significantly degrade the reach, in particular precluding the possibility to test
small multiplets.

\clearpage

Analogously to what we saw for CLIC, the bounds for multiplets with a mass roughly $1/4$ of the collider energy
suffer from a decreased sensitivity due to a suppression of the interference with the SM amplitude.
Collider runs at different energies can therefore provide complementary bounds for the small mass ranges.

Finally we notice that a $14\;$TeV muon collider could be useful to test several DM candidates, namely
the $(1,3,\epsilon)_{\rm DF}$, $(1,5,\epsilon)_{\rm DF}$ and $(1,7,\epsilon)_{\rm DF}$ multiplets, even for
mass values that saturate the DM relic abundance. It could also probe a significant fraction of the parameter space
of the $(1,5,0)_{\rm MF}$ multiplet. A $30\;$TeV muon collider could fully probe most of the DM candidates listed
in \Table{tableMDM}, provided enough integrated luminosity is collected.

\section{Comparison with direct searches}
\label{compDS}

Direct searches are particularly challenging 
if the lightest state within the EW multiplet is neutral, 
while if the lightest component is charged 
the current bounds are already quite strong. 
For instance, charged particles which are stable on the detector scale 
can be searched for by the longer time of flight through the detector and their 
anomalous energy loss. Current bounds are, 
depending on the quantum numbers, 
of the order of 300--900 GeV~\cite{DiLuzio:2015oha, Chatrchyan:2013oca, Khachatryan:2016sfv}. 

If the lightest state within the EW multiplet is neutral and stable, 
direct searches become more difficult and the bounds weaken.
Such states can be searched for in disappearing track searches or mono-X searches.\footnote{In non-minimal scenarios with extra states also soft lepton searches can be competitive \cite{Bharucha:2018pfu}.}~While a detailed assessment of direct searches is beyond the scope of this paper, 
we will shortly comment on the relevance of these searches for the EW multiplets considered in our analysis.

\underline{Disappearing tracks:}
The small mass splitting between the components of the EW multiplet 
implies a rather long lifetime of the next-to lightest charged component. 
This offers the possibility to search for them at colliders through disappearing charged tracks. In such searches a long-lived charged particle leaves a track in the innermost layers of the detector but not in the layers with higher radii, since it decays in the meanwhile into the neutral stable component of the EW multiplet and into a very soft pion, which cannot be reconstructed. While at Run-1 tracks needed to fly at least 29.9 cm in the ATLAS detector due to the requirement of at least one hit in the 
$b$-layer and three in the pixel detector \cite{Aad:2013yna} (similarly, also for the CMS Run-1 search \cite{CMS:2014gxa}), the installation of a new innermost layer in the ATLAS detector led to a significant improvement in the region of small lifetimes 
$(\sim 0.2 \text{ ns})$. For instance, at Run 1 EW triplets (``winos''), have been excluded up to 270 GeV \cite{Aad:2013yna}, Run-2 could exclude instead masses of 460 GeV \cite{Aaboud:2017mpt}. 
Remarkably, current bounds already outperformed projections on the reach of the HL-LHC before the detector upgrade for the wino \cite{Cirelli:2014dsa} and the Majorana 5-plet \cite{Ostdiek:2015aga}. 
For the latter, even though the lifetime is smaller 
than in the wino case (and hence the track length of the disappearing track is shorter, implying a less efficient search), 
one can benefit from the higher cross section. 
The projections derived in Ref.~\cite{Ostdiek:2015aga} for the Majorana 
5-plet were in fact stronger than for the wino.
\par
Finally, we mention that in Ref.~\cite{Fukuda:2017jmk} a new search strategy relying only on two hits 
(instead of four) in the innermost detectors was proposed. The benefit of this approach 
is a better sensitivity to states with shorter lifetimes (larger multiplets), 
requiring track lengths of $\gtrsim 5 \text{ cm}$. Following this strategy Ref.~\cite{Fukuda:2017jmk} found that winos can be excluded at the HL-LHC up to 1.2 TeV and EW doublets with $y=1/2$ (``higgsinos'') up to 550 GeV. 
In Ref.~\cite{Mahbubani:2017gjh} it was shown that performing a disappearing track search within the inner 10 cm of
the detector (hence modifying the detector set-up compared to ATLAS and CMS) would significantly improve the reach on higgsinos, allowing the FCC-100 to reach their thermal mass.
\par
\underline{Mono-X:}
Another way of testing EW multiplets in which the lightest component is electrically neutral and stable 
is by mono-X searches, where X stands for a high-energy jet, photon or a $W$/$Z$/Higgs boson. These searches require events with large missing transverse energy in association with hard SM radiation. 
While mono-X searches for DM-like states 
are more model independent, their reach is rather modest. 
The best sensitivity can usually be obtained 
from mono-jet searches, which are typically 
also more sensitive than vector boson fusion processes \cite{Cirelli:2014dsa}. 
Ref.~\cite{Han:2018wus} provides a comparison of the reach 
of disappearing track and mono-jet searches in the case of the wino and higgsino, 
both at the LHC and at future hadron colliders.  
It is found that the reach for winos (higgsinos) in monojet searches is 280 GeV (200 GeV) for the 
HL-LHC, 700 GeV (490 GeV)
at a 27 TeV collider and 2 TeV (1.4 TeV) at a 100 TeV collider, while the reach for disappearing track searches is roughly a factor of 3 larger for winos. Instead for higgsinos the disappearing track searches lead only to a slight improvement with respect to the reach of monojet searches. Monojet searches for larger multiplets were studied in Ref.~\cite{MDMmonojet}, where it is found 
that a MDM quintuplet can be constrained up to masses of roughly 700 GeV at the HL-LHC (with $\sqrt{s}=13 \text{ TeV}$) and up to 3.8 TeV at a  100 TeV collider with $L= 30\text{ ab}^{-1}$. 

For $e^+e^-$ machines we note that although direct searches at LEP almost allowed to saturate the kinematical 
reach for pair production, setting bounds on the mass of electroweakly charged states of the order 
of $\sqrt{s} / 2 - \epsilon$, with $\epsilon \approx 10 \% \, \sqrt{s}$ \cite{Abbiendi:2002vz}, this will not necessarily be the case at high-energy lepton colliders like CLIC. The reason is that some SM backgrounds, 
like the vector-boson-fusion production of a $Z$ boson decaying into neutrinos or 
the real emission from Bhabha scattering, 
become more relevant at high energies.\footnote{We thank Roberto Franceschini 
for clarifications regarding this point.} Hence, indirect probes of EW states based on 
precision measurement become 
relevant also at energies above the pair-production threshold. 

In summary, while direct searches lead to stronger bounds on small multiplets 
(doublets and triplets), for larger multiplicities (e.g.~5-plets) 
precision measurements seem to surpass the projected sensitivities from direct searches.  
A detailed comparison is however beyond the scope of this paper, since 
a dedicated analysis of direct searches for EW multiplets with $n>3$ 
after detector upgrades is not available yet in the literature. 

\section{Conclusions}\label{sec:conclusion}

In this paper we investigated the possibility of probing EW multiplets by precision measurements at the LHC and future high-energy hadron and lepton colliders.
We first focused on the Drell-Yan production of $\ell^+ \ell^-$ and $\ell \nu$ at hadron colliders, which are modified in the presence of new EW multiplets due to one-loop corrections of the gauge boson propagators. 
\par
We compared in detail neutral and charged Drell-Yan production and
showed that the charged current channel actually leads to much stronger bounds
than the previously considered neutral current channel \cite{Matsumoto:2017vfu}.
While we find that an ${\rm SU}(2)_L$ doublet can neither be probed at the LHC nor at future
hadron colliders since the modifications of the SM processes remain below
the systematic uncertainty, for larger multiplets the prospects are much more 
promising. For instance, we find that a future 100 TeV collider
can fully probe the DM hypothesis in the case of a $(1,5,\epsilon)_{\text{DF}}$ state. 
For other EW multiplets the thermal mass cannot be reached but we can 
still constrain part of the allowed parameter space. 
The sensitivity on EW multiplets is just below
the reach of disappearing track searches, 
although indirect EW probes are more model-independent 
(e.g.~they do not depend on kinematical features like the lifetime of the 
inter-multiplet components). 
A precision measurement
of the Drell-Yan process hence provides an alternative, 
model-independent probe of
EW multiplets. By reducing systematic uncertainties, for instance via a
better knowledge of PDFs, the reach of our method increases significantly. 
\par 
Furthermore, we considered the possibility of constraining EW multiplets
via the process $\ell^+ \ell^-\to f\bar{f}$ at a future $e^+e^-$ or $\mu^+\mu^-$ collider.  
The fundamental difference 
compared to hadron colliders is that the centre-of-mass energy is fixed. 
Exclusion bounds thus strongly depend on the shape of the form factors
of the new physics contribution to the gauge boson self-energies.
Since the sensitivity is not a monotonously decreasing function with the mass of the new
multiplet, different centre-of-mass energies can probe 
complementary mass ranges.
Different stages for the centre-of-mass energy are hence 
potentially helpful for probing new EW particles at lepton colliders. 
Employing the energy and luminosity benchmarks of CLIC we find that 
it is possible to reach the thermal mass of a 
$(1,3,\epsilon)_{\rm DF}$ multiplet, 
while for the recently revived option of a high-energy and high-luminosity
muon collider 
also the $(1,5,\epsilon)_{\rm DF}$ and $(1,7,\epsilon)_{\rm DF}$ thermal masses can be tested.

\section*{Acknowledgments}

We are indebted with Roberto Franceschini for triggering this work and for 
many useful discussions. 
We also acknowledge helpful discussions with 
Jorge De Blas, Eugenio Del Nobile, Jonas Lindert, 
Marco Nardecchia, Paolo Panci, Thomas Rauh, Filippo Sala, 
Michael Spannowsky and Andrea Wulzer. 
The work of LDL was supported by the ERC grant NEO-NAT.
RG was partially supported by a COFUND/Durham Junior Research Fellowship under the EU grant number 609412.

\section*{Note added} 
While completing this work a preprint appeared on the arXiv \cite{EWIMPs100} 
where bounds on EW multiplets for the neutral Drell-Yan channel at a future 100 TeV collider were discussed.
As we showed however, the charged current channel leads to stronger bounds than the neutral one also for a future 100 TeV collider.

\appendix

\section{Perturbativity}
\label{App:pert}

Large representations $\chi \sim (1,n,y)$ will eventually lead to a breakdown of the perturbative expansion. 
In principle, to assess the perturbative stability of the bounds extracted in this paper 
one should consider the two-loop corrections to the EW self-energies \cite{Djouadi:1993ss},  
which is however beyond our scopes.  
Nevertheless, we can perform a simple estimate of the domain of perturbativity by requiring that the 
two-loop 
correction to the beta function of the gauge coupling $g_i$ (either $g$ or $g'$) does not overcome 
a certain fraction $f$ of 
the one-loop contribution.\footnote{An alternative criterium 
is to require $|\beta^{(1)}_{g_i} / g_i| < 1$ \cite{Goertz:2015nkp}, 
where $\beta^{(1)}_{g_i}$ is the one-loop beta function.} 
Focussing on the part of the two-loop correction that dominates either for large $n$ or $y$ 
(see Ref.~\cite{Machacek:1983tz} 
for complete formulae), we find: 
\begin{align}
\left(\frac{g_i}{4\pi}\right)^2 4 C_2 (F_i) S_2(F_i) &< f \frac{4}{3} S_2(F_i)  \qquad \text{(fermions)} \, , \\
\left(\frac{g_i}{4\pi}\right)^2 4 C_2 (S_i) S_2(S_i) &< f \frac{1}{3} S_2(S_i)  \qquad \text{(scalars)} \, .
\end{align}
Here, $S_2$ and $C_2$ denote respectively the Dynkin index 
and the quadratic Casimir of a given fermionic ($F$) or scalar ($S$) representation. 
For ${\rm SU}(2)_L$ one has $S_2 = n (n^2-1)/12$ and $C_2 = (n^2-1)/4$, while for ${\rm U}(1)_y$ 
one has $S_2 = C_2 = y^2$. Hence, in terms of $n$ and $y$ we have  
\begin{align}
n &< \sqrt{f \left(\frac{4 \pi}{g}\right)^2 \frac{4}{3} +1} 
\qquad \text{and} \qquad
y < \sqrt{\frac{f}{3}} \left(\frac{4 \pi}{g'}\right)
\qquad \text{(fermions)} \, , \\
n &< \sqrt{f \left(\frac{4 \pi}{g}\right)^2 \frac{1}{3} +1} 
\qquad \text{and} \qquad
y < \sqrt{\frac{f}{12}} \left(\frac{4 \pi}{g'}\right)
\qquad \text{(scalars)} \, .
\end{align}
Taking as reference values $g = 0.65$, $g' = 0.36$ and $f = 50\%$, we find 
$n < 16 \, (8)$ and $y < 14 \, (7)$ for fermions (scalars). 
Note, however, that in the case of Dirac fermions 
already $n \gtrsim 9$ can develop Landau poles within one order of magnitude 
from the mass scale of $\chi$ (cf.~Table 9 in Ref.~\cite{DiLuzio:2015oha}). 
For this reason we cut our plots below $n=10$.

\section{Additional results for future hadron colliders}\label{app:additional_results}
\label{App:additional}

In this appendix we collect some additional results regarding the expected bounds at high-energy hadron colliders.

\begin{figure}[t]
\includegraphics[width=0.48 \textwidth]{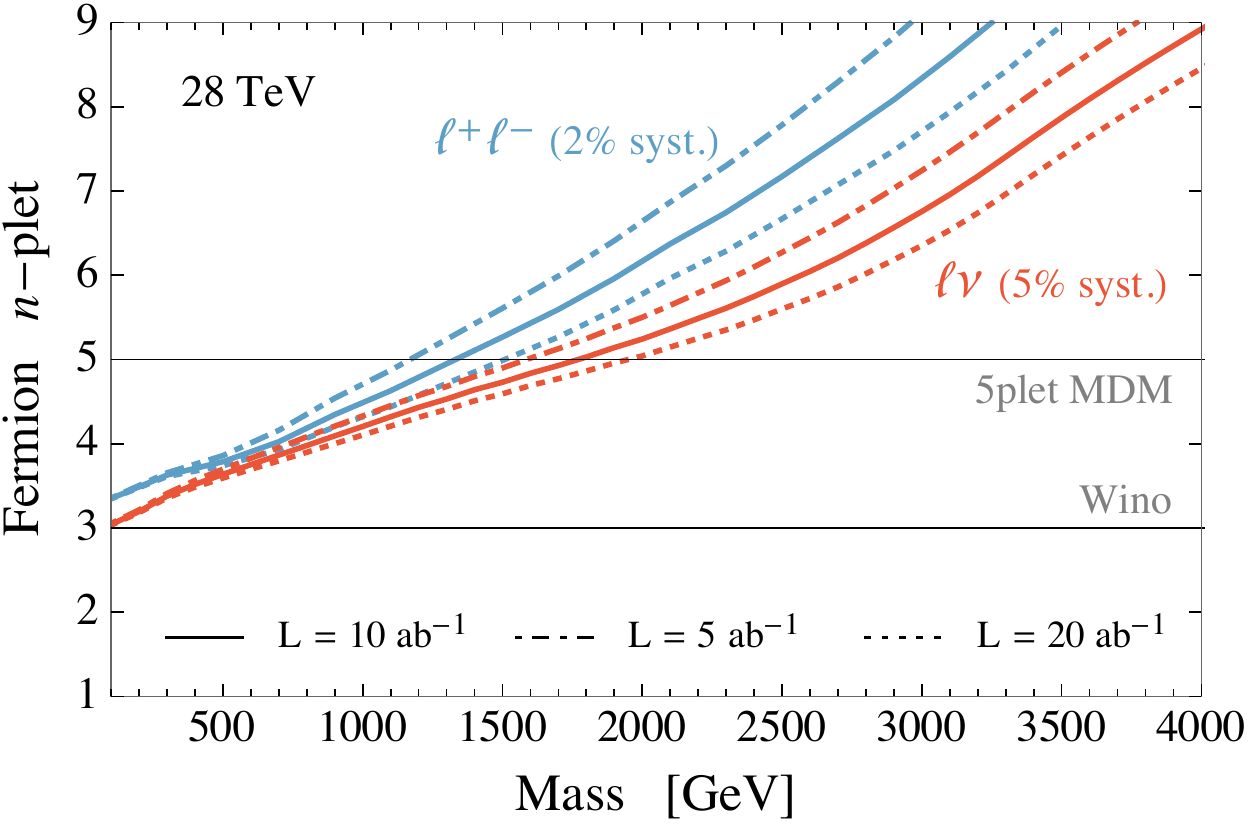}
\hfill
\includegraphics[width=0.48 \textwidth]{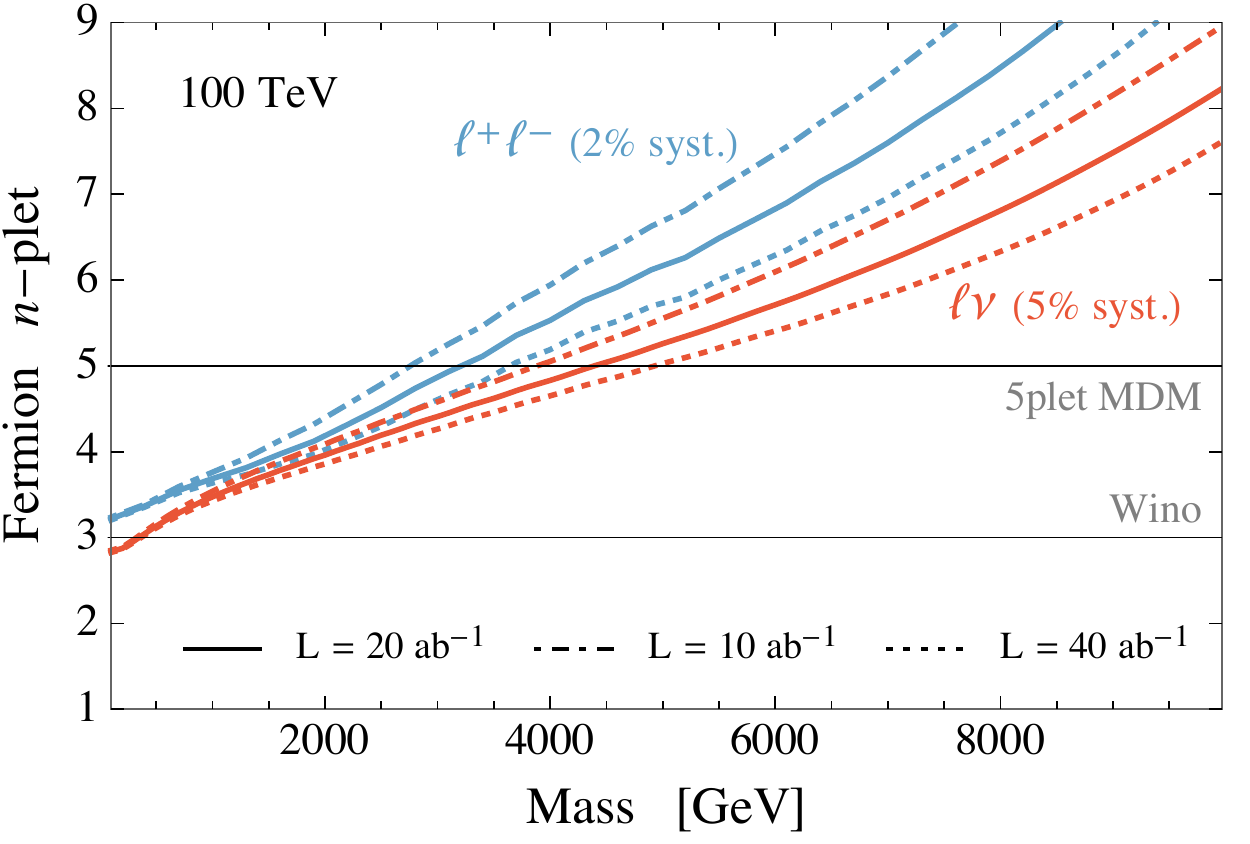}
\caption{\it Dependence of the $95\%$ CL exclusion bounds on the integrated luminosity at the HE-LHC (left panel)
and at the FCC-100 (right panel). The red and blue lines correspond to the bounds from the $\ell\nu$ and
$\ell^+\ell^-$ final states.}\label{fig:luminosity}
\end{figure}

In \fig{fig:luminosity} we show how the expected bounds for Majorana fermion multiplets depend on the
integrated luminosity at the HE-LHC (left panel) and FCC-100 (right panel). The solid lines correspond to the
benchmark luminosities used in the main text, while the dotted and dot-dashed lines are obtained by doubling and
halving the integrated luminosity respectively. One can see that the impact of a luminosity change is more important
for high masses (or equivalently larger multiplets), where the mass bounds change by roughly $\pm 10\%$ by
doubling/halving the luminosity. This behavior is due to the fact that for low masses low luminosities are already enough
to give a statistical error below the systematic ones, whereas for larger masses statistical errors still play a relevant role
for the benchmark luminosity we considered.

\begin{figure}[t]
\includegraphics[width=0.48 \textwidth]{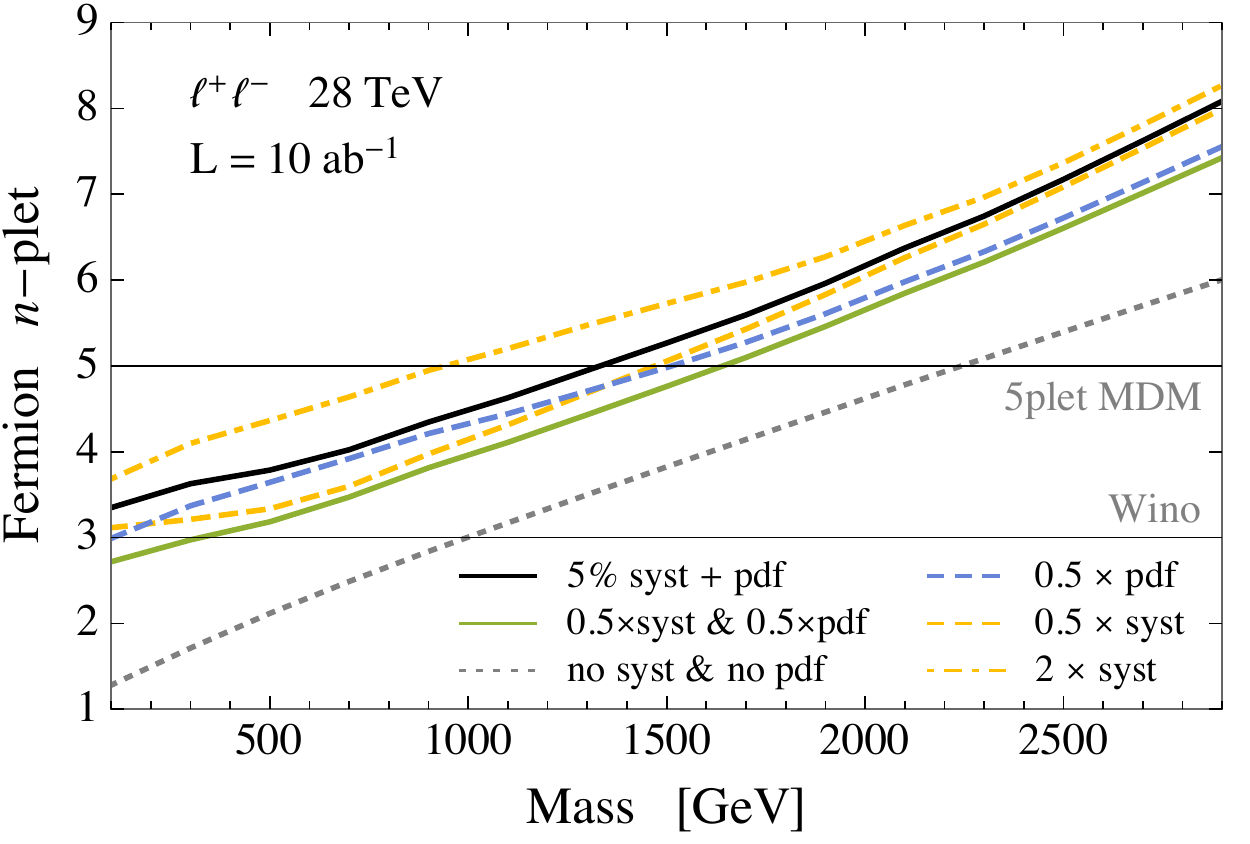}
\hfill
\includegraphics[width=0.48 \textwidth]{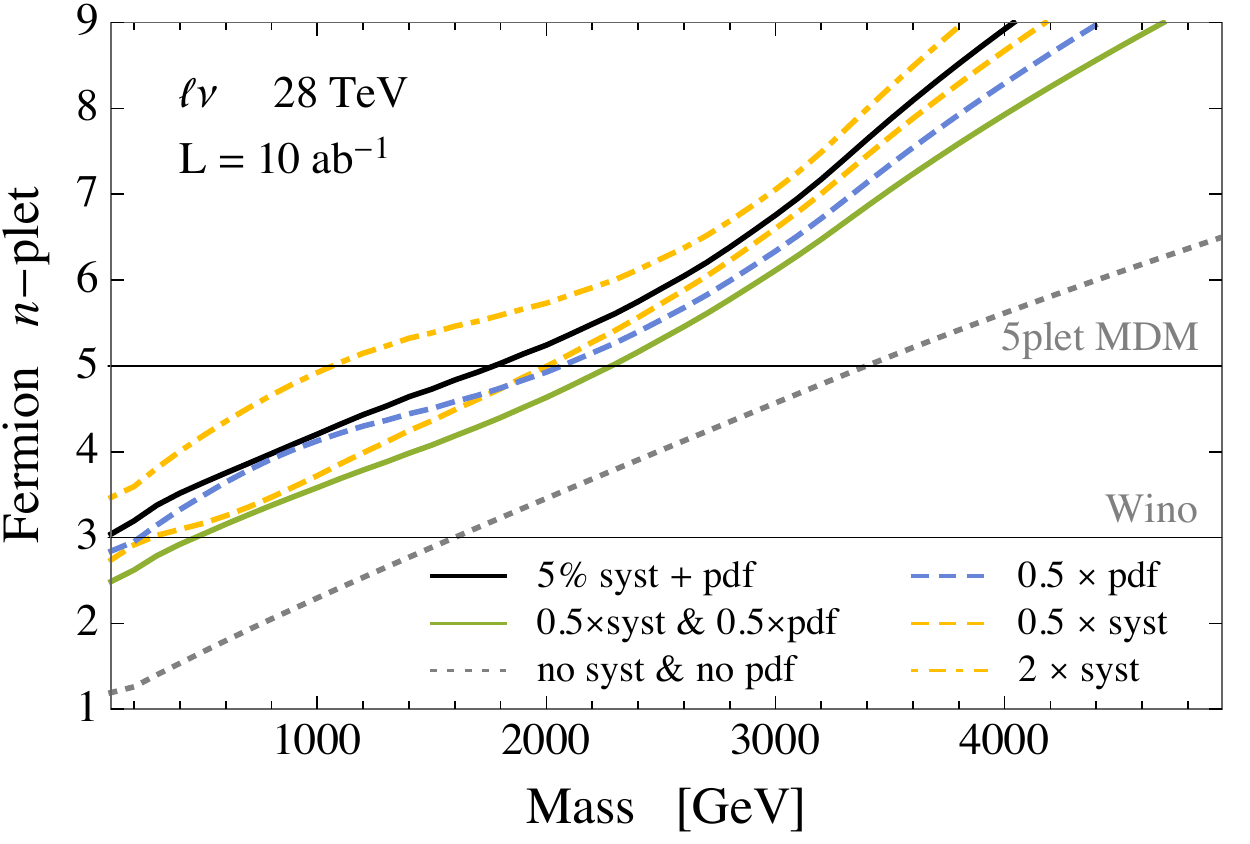}
\caption{\it Dependence of the $95\%$ CL exclusion bounds on Majorana fermion multiplets at the HE-LHC
on the PDF uncertainties and on the additional systematic errors.}\label{fig:syst_unc_28TeV}
\end{figure}

\begin{figure}[t]
\includegraphics[width=0.48 \textwidth]{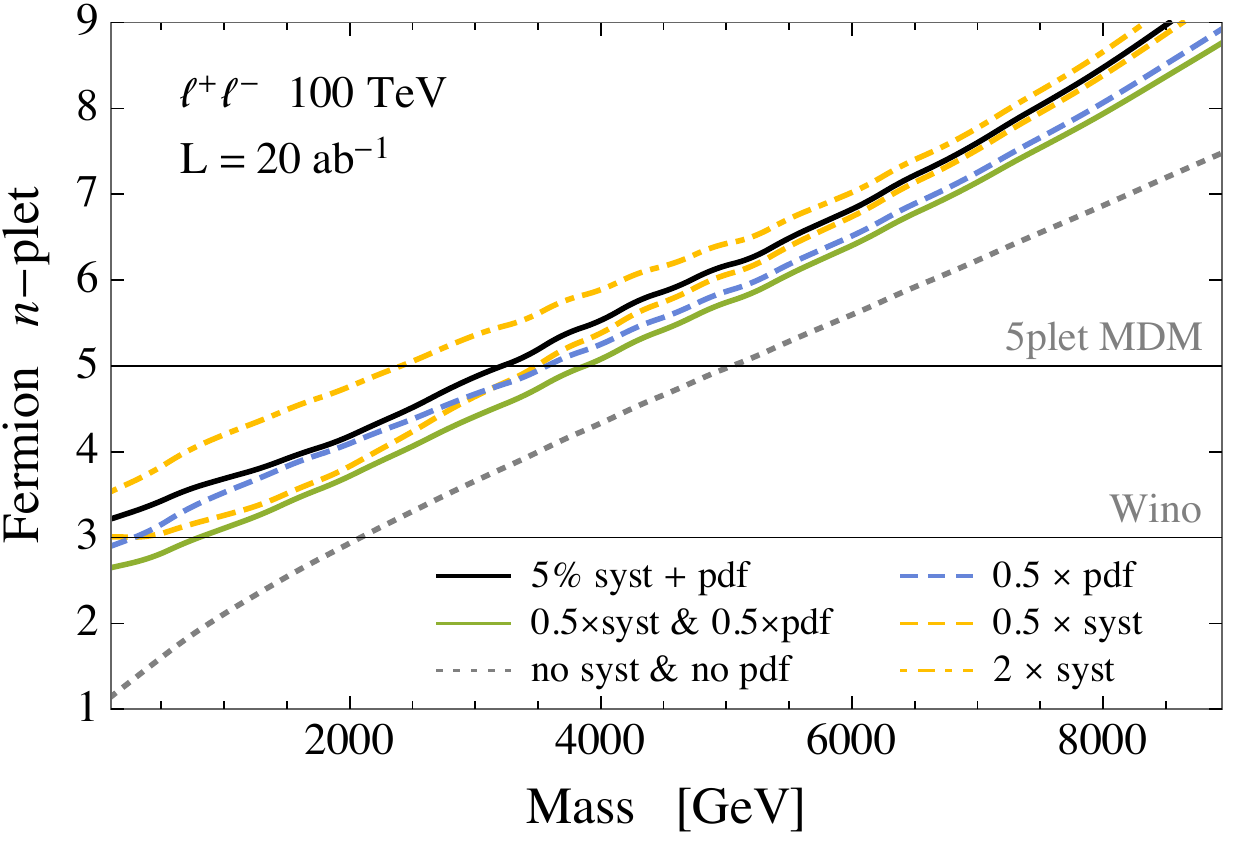}
\hfill
\includegraphics[width=0.48 \textwidth]{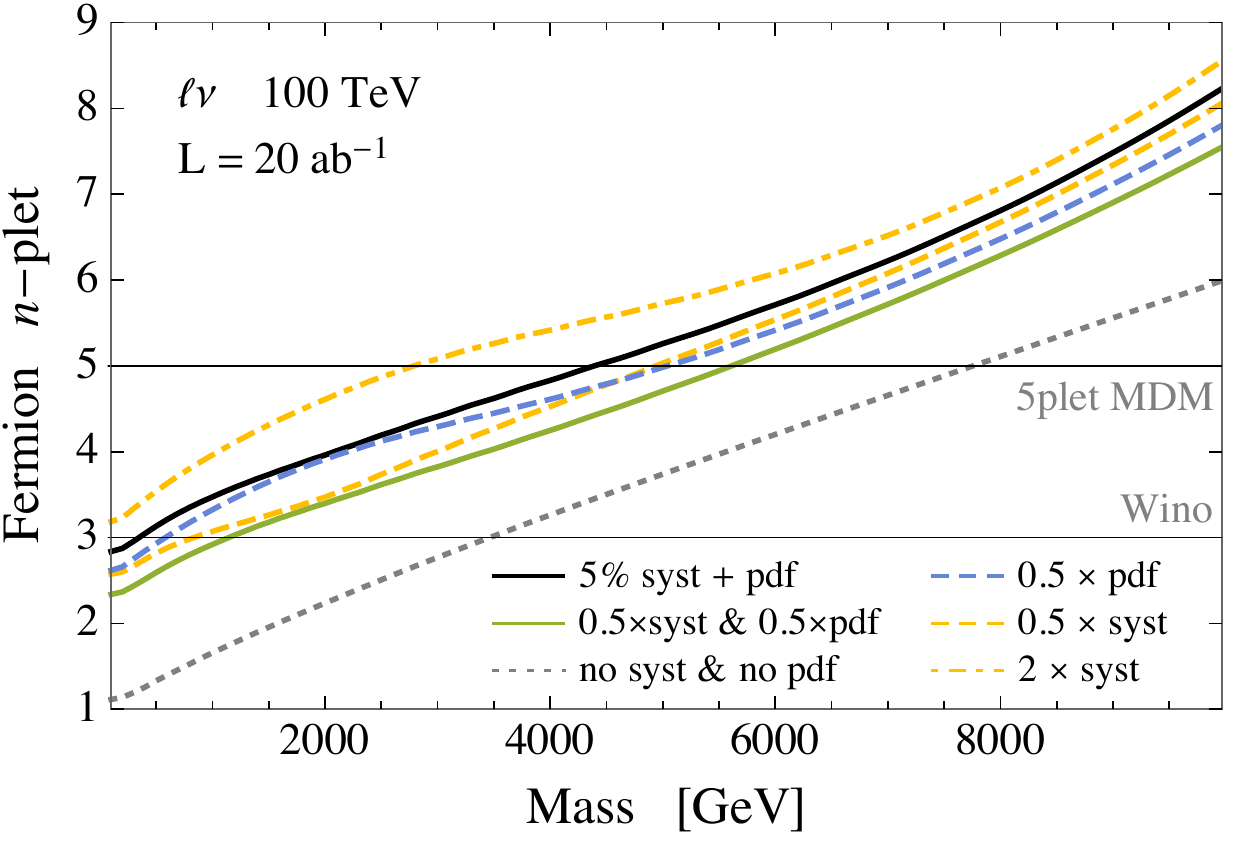}
\caption{\it Dependence of the $95\%$ CL exclusion bounds on Majorana fermion multiplets at the FCC-100
on the PDF uncertainties and on the additional systematic errors.}\label{fig:syst_unc_100TeV}
\end{figure}

In Figs.~\ref{fig:syst_unc_28TeV} and \ref{fig:syst_unc_100TeV} we show how the expected bounds change by varying
our assumptions about PDF and additional systematic errors. The behavior is qualitatively similar to the one we
discussed for the HL-LHC in \sect{sec:syst_unc}. PDF uncertainties tend to dominate for higher masses
(above $\sim 2\;$TeV for HE-LHC and above $\sim 4\;$TeV for FCC-100), while the additional systematics
are more relevant for lower masses. A reduction by $50\%$ of the systematics allows for a $\sim15\%$ improvement
in the bounds at high masses and a nearly $100\%$ improvement for low masses. In particular such an improvement
would allow hadron colliders to test Majorana fermion triplets, which can not be probed with the benchmark systematics we
assumed in the main text.

\clearpage

\bibliographystyle{utphys.bst}
\bibliography{bibliography}

\end{document}